\documentclass[preprint, aps, prd]{revtex4-1}

\def\pd{\partial}
\def\mc{\mathcal}

\usepackage[dvips]{graphicx}
\usepackage{amssymb}
\usepackage{amssymb,amsmath}
\usepackage{graphicx}
\usepackage{dsfont}
\usepackage{caption}
\usepackage{subcaption}
\makeatother
\begin{document}

\title{Supersymmetric solutions from $N=5$ gauged supergravity}

\author{Parinya Karndumri} \author{Chawakorn Maneerat} \email[REVTeX Support:
]{parinya.ka@hotmail.com and chawakorn.manee@gmail.com} 
\affiliation{String Theory and
Supergravity Group, Department of Physics, Faculty of Science,
Chulalongkorn University, 254 Phayathai Road, Pathumwan, Bangkok
10330, Thailand}

\date{\today}
\begin{abstract}
We study a large class of supersymmetric solutions in four-dimensional $N=5$ gauged supergravity with $SO(5)$ gauge group. There is only one $N=5$ supersymmetric $AdS_4$ vacuum preserving the full $SO(5)$ symmetry dual to an $N=5$ SCFT in three dimensions. We give a number of domain walls interpolating between this $AdS_4$ fixed point and singular geometries in the IR with $SO(4)$ and $SO(3)$ symmetries. These solutions describe RG flows from the $N=5$ SCFT to non-conformal field theories driven by mass deformations. The $SO(4)$ solutions are precisely in agreement with the previously known mass deformations within the dual $N=5$ SCFT. We also find supersymmetric Janus solutions describing two-dimensional conformal defects in the $N=5$ SCFT with $N=(4,1)$ and $N=(1,1)$ supersymmetries on the defects. Finally, we study supersymmetric solutions of the form $AdS_2\times \Sigma^2$, with $\Sigma^2=S^2,H^2$ being a Riemann surface, corresponding to near horizon geometries of $AdS_4$ black holes. We consider both magnetic and dyonic solutions and find that there exists a class of magnetic $AdS_2\times H^2$ solutions with $SO(2)$ symmetry. It is rather remarkable that a complete analytic solution interpolating between $AdS_4$ and $AdS_2\times H^2$ with a running scalar can be obtained. The solution corresponds to a twisted compactification of $N=5$ SCFT to superconformal quantum mechanics. We also show that no purely magnetic or dyonic black holes with $AdS_2\times \Sigma^2$ horizon from $SO(2)\times SO(2)$ twist exist in $N=5$, $SO(5)$ gauged supergravity.
\end{abstract}
\maketitle
%%%%%%%%%%%%%%%%%%%%%%%%%%%%%%%%%%%%%%%%%%%%%%%%%%%%%%%%%%%%%%%%%%%%%%%%%%%%%%%%%%%%%%%%%%%%%%%%%%%%%%%%%%%%%%%%%%%%%%%%%%%%%%%%%%%%%%%%%
\section{Introduction}
Over the past twenty years, the AdS/CFT correspondence, originally proposed in \cite{maldacena} see also \cite{Gubser_AdS_CFT,Witten_AdS_CFT}, has provided holographic descriptions of various strongly coupled systems ranging from (non) conformal field theories, conformal defects, $AdS$-black holes and condensed matter physics systems. Although the complete AdS/CFT duality is achieved only in the context of string/M-theory, a large number of remarkable results has been obtained from solutions of lower-dimensional gauged supergravities.
\\
\indent In many cases, the gauged supergravities under consideration are known to be consistently embedded in ten- or eleven-dimensional supergravities which are low energy effective theories of string/M-theory. The resulting solutions can accordingly be uplifted to string/M-theory and can be interpreted as D- and M-brane configurations. Solutions of gauged supergravities with presently unknown higher-dimensional origins are also interesting in the sense that they can provide a bottom-up approach to the AdS/CFT duality and still give some insight to the dynamics of the dual field theories at strong coupling limits. These make studying solutions of gauged supergravities in various space-time dimensions and different numbers of supersymmetries worth considering.
\\
\indent Most of the previous studies concern with finding a particular class of solutions that preserve some amount of supersymmetry. These supersymmetric or BPS solutions play an important role in different aspects of the AdS/CFT correspondence. Gauged supergravities including possible massive deformations are known to exist in dimensions from two to ten. Among these theories, four-dimensional guaged supergravities are of particular interest since they give rise to holographic duals of three-dimensional superconformal field theories (SCFTs) and possible deformations thereof. These SCFTs describe low energy dynamics of the world-volume theory on M2-branes which are fundamental objects in M-theory. The SCFTs in three dimensions take the form of Chern-Simons-Matter (CSM) theories since the usual gauge theories with Yang-Mills gauge kinetic terms are not conformal. Up to now, many of these SCFTs with different numbers of supersymmetries have been constructed, see \cite{Bena,BL1,BL2,BL3,Gustavsson,Basu_Harvey,Schwarz_3D_SCFT,ABJM,ABJ,N5_6_3D_SCFT,ST,BL4,LP,Origin_3_algebra,
Eric_3D_supercon,Chen,Eric_M2_brane,Eric_Mbrane_SUGRA,MFM,BB,Honda} for an incomplete list.
\\
\indent In this paper, we are interested in supersymmetric solutions of $N=5$ gauged supergravity with $SO(5)$ gauge group constructed long ago in \cite{de_Wit_N5}. According to the AdS/CFT duality, these solutions could describe various aspects of strongly coupled $N=5$ SCFTs in three dimensions. There are ten scalars described by $SU(5,1)/U(5)$ coset. The scalar potential of this gauged supergravity has been analyzed in \cite{Warner_Potential}. There is only one supersymmetric $AdS_4$ vacuum preserving the full $N=5$ supersymmetry with unbroken $SO(5)$ symmetry, see also a general discussion in \cite{Severin_Maximal_AdS}. According to the AdS/CFT duality, this $AdS_4$ critical point is dual to an $N=5$ SCFT in three dimensions. There is also another non-supersymmetric $AdS_4$ vacuum with unbroken $SO(3)$ gauge symmetry. This critical point is perturbatively stable as pointed out in \cite{Warner_Fisch} and has been extensively studied in the context of holographic superconductors in \cite{Bobev_holo_supercon}. To the best of our knowledge, no supersymmetric solutions of $N=5$ gauged supergravity have previously been considered. The present work will hopefully fill this gap in the existing literature.
\\
\indent We will look for various types of supersymmetric solutions of the aforementioned $N=5$ gauged supergravity. Firstly, we will study supersymmetric domain walls interpolating between the supersymmetric $AdS_4$ vacuum and singular geometries. These solutions describe holographic RG flows from the dual $N=5$ SCFT in the UV to non-conformal field theories in the IR obtained from mass deformations of the $N=5$ SCFT. Similar solutions have been extensively studied in $N=8$ and $N=2$ four-dimensional gauged supergravities, see for example \cite{Warner_membrane_flow,Warner_M_F_theory_flow,Warner_higher_Dflow,Flow_in_N8_4D,4D_G2_flow,
Warner_M2_flow,Guarino_BPS_DW,Elec_mag_flows,Yi_4D_flow}. In addition, solutions in gauged supergravities with $N=3,4$ supersymmetries have been considered recently in \cite{N3_SU2_SU3,N3_4D_gauging,tri-sasakian-flow,orbifold_flow,4D_N4_flows}.
\\
\indent We will also find Janus solutions described by $AdS_3$-sliced domain walls interpolating between asymptotically $AdS_4$ spaces. These solutions are holographically dual to two-dimensional conformal defects within the $N=5$ SCFT that break the superconformal symmetry in the three-dimensional bulk to a smaller superconformal symmetry on the two-dimensional surfaces. Supersymmetric Janus solutions in other four-dimensional gauged supergravities have previously been studied in \cite{tri-sasakian-flow,orbifold_flow,warner_Janus,N3_Janus,Minwoo_4DN8_Janus,Kim_Janus}.
\\
\indent We finally look for solutions interpolating between the supersymmetric $AdS_4$ and $AdS_2\times \Sigma^2$ geometries with $\Sigma^2$ being a Riemann surface. These solutions describe supersymmetric black holes in an asymptotically $AdS_4$ space. Solutions of this type in other gauged supergravities can be found in \cite{BH_M_theory1,BH_M_theory2,AdS4_BH1,AdS4_BH2,AdS4_BH3,AdS4_BH4,AdS4_BH5,
Guarino_AdS2_1,Guarino_AdS2_2,Trisasakian_AdS2}. In the dual field theory, the solutions are dual to RG flows from the $N=5$ SCFT to another SCFT in one dimension or superconformal quantum mechanics. The latter is obtained from the former via twisted compactifications on $\Sigma^2$. This type of solutions plays an important role in microscopic computation of black hole entropy in asymptotically $AdS_4$ space, see for example \cite{Zaffaroni_BH_entropy,BH_entropy_benini,BH_entropy_Passias}. We finally note that the $N=5$ gauged supergravity with $SO(5)$ gauge group is a subsector of the $SO(8)$ $N=8$ gauged supergravity arising from a consistent truncation of M-theory on $S^7$. Solutions given in this paper are also solutions of the maximal gauged supergravity and can accordingly be embedded in M-theory. On the other hand, it has been pointed out in \cite{N5_6_3D_SCFT} that the $N=5$ SCFT with $SO(2N)\times Sp(2N)$ gauge group can be obtained from the ABJM theory with $U(2N)\times U(2N)$ gauge group via orientifolds. We then expect this $N=5$ SCFT to be dual to the $N=5$ $AdS_4$ vacuum.   
\\
\indent The paper is organized as follows. In section \ref{N5_SUGRA},
we review the construction of four-dimensional $N=5$ gauged supergravity with $SO(5)$ gauge group. In section \ref{RG_flow}, we will look for supersymmetric $AdS_4$ vacua and domain wall solutions describing RG flows in the dual $N=5$ SCFTs in three-dimensions. We then study supersymmetric Janus solutions in section \ref{Janus_N5}, and finally look for possible supersymmetric $AdS_4$ black holes for both magnetic and dyonic solutions in section \ref{AdS4_BH}. Conclusions and comments on the results are given in section \ref{conclusion}.

%%%%%%%%%%%%%%%%%%%%%%%%%%%%%%%%%%%%%%%%%%%%%%%%%%%%%%%%%%%%%%%%%%%%%%%%%%%%%%%%%%%%%%%%%%%%%%%%%%%%%%%%%%%%%%%%%%%%%%%%%%%%%%%%%%%%%%%%%
\section{$N=5$ gauged supergravity with $SO(5)$ gauge group}\label{N5_SUGRA}
We begin with a review of $N=5$ gauged supergravity constructed in \cite{de_Wit_N5}. We also follow most of the convention in \cite{de_Wit_N5}, see more detail on the convention in \cite{de_Wit_N8}, but with some slightly modified notations to match with the modern notation reviewed in \cite{Mario_Physics_Rep}. $N=5$ supersymmetry does not allow for any matter multiplets, so the only supermultiplet in $N=5$ supergravity is the gravity multiplet with the following field content
\begin{equation}
(e^a_\mu,\psi^i_\mu,A^{ij}_\mu, \chi^{ijk},\chi,\phi^i).
\end{equation}
The component fields correspond to the graviton $e^a_\mu$, five gravitini $\psi^i_\mu$, ten vectors $A^{ij}_\mu=-A^{ji}_\mu$, eleven spin-$\frac{1}{2}$ fields $\chi^{ijk}=\chi^{[ijk]}$ and $\chi$ together with five complex scalars $\phi^i$.
\\
\indent Space-time and tangent space indices are denoted by $\mu,\nu,\ldots =0,1,2,3$ and $a,b,\ldots =0,1,2,3$, respectively. The $N=5$ supergravity admits global $SU(5,1)$ and local composite $U(5)\sim SU(5)\times U(1)$ symmetries. The latter is the R-symmetry for $N=5$ supersymmetry. Indices $i,j,k,\ldots=1,2,\ldots, 5$ denote fundamental representation of $SU(5)$. The ten scalars are described by $SU(5,1)/U(5)$ coset manifold with the coset representative
\begin{equation}
{\Sigma^A}_B=\left(\begin{array}{c|c}
                                           {\delta^i}_j-e_2\phi^i\phi_j & e_1\phi^i \\ \hline
                                           e_1\phi_j & e_1 \\
                                         \end{array}
                                       \right)
 \end{equation}
with $A,B=1,2,\ldots, 6$ being indices of $SU(5,1)$ fundamental representation. The quantities $e_1$ and $e_2$ are defined by
\begin{equation}
e_1=\frac{1}{\sqrt{1-|\phi|^2}}\qquad \textrm{and}\qquad e_2=\frac{1}{|\phi|^2}\left(1-\frac{1}{\sqrt{1-|\phi|^2}}\right)
\end{equation}
with $|\phi|^2=\phi^i\phi_i=\phi^i(\phi^i)^*$. It should be noted that the notaion $\phi_i$ denotes the complex conjugate of $\phi^i$. In addition, in this parametrization, $\phi^i$ satisfy the condition $|\phi|^2<1$. Being an element of $SU(5,1)$, ${\Sigma^A}_B$ satisfies the following identity
\begin{equation}
\Sigma^{-1}=\eta \Sigma^\dagger \eta
\end{equation}
in which $\eta=\textrm{diag}(1,1,1,1,1,-1)$ is the $SU(5,1)$ invariant tensor.
\\
\indent The ten vector fields $A^{ij}_\mu$ can be used to gauge $SO(5)\subset SU(5)\subset SU(5,1)$ symmetry resulting in $N=5$ gauged supergravity with $SO(5)$ gauge group. The corresponding bosonic Lagrangian is given by
\begin{eqnarray}
e^{-1}\mc{L}&=&\frac{1}{2}R-\frac{1}{2}P^i_\mu P^{\mu}_i-\frac{1}{8}\left[(2S^{ij,kl}-\delta^{ik}\delta^{jl})F^+_{\mu\nu ij}F^{+\mu\nu}_{kl} \right.\nonumber \\
& &\left. +(2S_{ij,kl}-\delta_{ik}\delta_{jl})F^{-ij}_{\mu\nu}F^{-\mu\nu kl}\right]-V\label{L_N5}
\end{eqnarray}
with the $10\times 10$ matrix $S_{ij,kl}=(S^{ij,kl})^*$. As usual, the vielbein $P^i_\mu$ and $SU(5)\times U(1)$ composite connection ${{Q_\mu}^i}_j$ on the scalar manifold are obtained from the relation
\begin{equation}
\Sigma^{-1}D_\mu \Sigma=\left(\begin{array}{c|c}
                                           \frac{1}{2}{{Q_\mu}^i}_j-\frac{1}{6}{\delta^i}_j{{Q_\mu}^k}_k& -\frac{1}{\sqrt{2}}P^i_\mu \\ \hline
                                           -\frac{1}{\sqrt{2}}P_{\mu j} & \frac{1}{3}{{Q_\mu}^k}_k \\
                                         \end{array}
                                       \right).
 \end{equation}
Explicitly, we can write
\begin{eqnarray}
P^i_\mu&=&-\sqrt{2}e_1({\delta^i}_j-e_2\phi^i\phi_j)D_\mu \phi^j,\\
{{Q_\mu}^i}_j&=&2e_2\phi^i\overleftrightarrow{D}_\mu \phi_j+\frac{1}{2}(e_1^2{\delta^i}_j-2e_2^2\phi^i\phi_j)\phi^k\overleftrightarrow{D}_\mu \phi_k
\end{eqnarray}
with the gauge covariant derivative given by
\begin{equation}
D_\mu \phi_i=\pd_\mu \phi_i-gA^{ij}_\mu \phi_j
\end{equation}
and $D_\mu \phi^i=(D_\mu \phi_i)^*$. We also note the properties of ${{Q_\mu}^i}_j$
\begin{equation}
{{Q_\mu}^i}_j=-{Q_{\mu j}}^i=-({{Q_\mu}^i}_j)^*\, .
\end{equation}
\indent We now come to the gauge field part. The (anti) self-dual field strength tensors are defined as
\begin{equation}
F^{+}_{\mu\nu ij}=\frac{1}{2}\left(F_{\mu\nu}^{ij}+ \frac{i}{2}\epsilon_{\mu\nu\rho\sigma}F^{ij\rho\sigma}\right)\qquad \textrm{and}\qquad F^{-ij}_{\mu\nu}=\frac{1}{2}\left(F_{\mu\nu}^{ij}- \frac{i}{2}\epsilon_{\mu\nu\rho\sigma}F^{ij\rho\sigma}\right)
 \end{equation}
with the gauge field strengths defined by
\begin{equation}
F^{ij}_{\mu\nu}=2\pd_{[\mu}A_{\nu]}^{ij}-2gA^{ik}_{[\mu}A^{kj}_{\nu]}\, .
\end{equation}
We also note the convention $(F^+_{\mu\nu ij})^*=F^{-\ij}_{\mu\nu}$.
\\
\indent The matrix $S^{ij,kl}$ is defined by the relation
\begin{equation}
\left(\delta^{ij}_{kl}+\frac{1}{2}\epsilon^{ijklp}\phi_p\right)S^{kl,mn}=\delta^{ij}_{mn}
\end{equation}
with $\delta^{ij}_{kl}=\frac{1}{2}(\delta^i_k\delta^j_l-\delta^i_l\delta^j_k)$. The explicit form of $S^{ij,kl}$ can be found in \cite{Ferrara_N5_BH} in which black hole attractors in ungauged $N=5$ supergravity have been studied. In our notation, this matrix reads
\begin{equation}
S^{ij,kl}=\frac{1}{1-(\phi_n)^2}\left[\delta^{ij}_{kl}-\frac{1}{2}\epsilon^{ijklm}\phi_m-2\delta_{[i[k}\phi_{l]}\phi_{j]}\right].
\end{equation}
\indent Finally, the scalar potential is given by
\begin{equation}
V=-g^2\left[2+4e^2_1-\frac{1}{2}e_1^4\left(|\phi|^2-(\phi_i)^2(\phi^j)^2\right)\right]
\end{equation}
with the notation $(\phi_i)^2=\phi_i\phi_i$.
\\
\indent We also need supersymmetry transformation rules for fermions. In the chiral notation, the fermionic fields are subject to the chirality projection
\begin{equation}
\gamma_5\psi^i_\mu =\psi^i_\mu, \qquad \gamma_5\chi=-\chi,\qquad \gamma_5\chi^{ijk}=\chi^{ijk}
\end{equation}
with
\begin{equation}
\gamma_5\psi_{i\mu} =-\psi_{i\mu}, \qquad \gamma_5\chi_{ijk}=-\chi_{ijk}\, .
\end{equation}
The corresponding supersymmetry transformations read
\begin{eqnarray}
\delta\psi_{\mu i}&=&2D_\mu \epsilon_i-{{Q_\mu}^j}_i\epsilon_j-\frac{1}{2\sqrt{2}}\gamma^{\nu\rho}\gamma_\mu G^+_{\nu\rho kl}{C_{ij}}^{kl}\epsilon^j+\sqrt{2}g\gamma_\mu S_{ij}\epsilon^j,\\
\delta \chi_{ijk}&=&-\epsilon_{ijklm}P^m_\mu \gamma^\mu \epsilon^l+\frac{3}{2}G^+_{\mu\nu rs}\gamma^{\mu\nu}{C_{[ij}}^{rs}\epsilon_{k]}-2g{N^l}_{ijk}\epsilon_l,\label{d_chi}\\
\delta\chi&=&P_{\mu i}\gamma^\mu \epsilon^i-2gN^i\epsilon_i
\end{eqnarray}
in which the fermion-shift matrices are defined by
\begin{eqnarray}
S^{ij}&=&e_1\delta^{ij}+\frac{1}{2}e_2^2\left[|\phi|^2(\phi^i\phi_j+\phi_i\phi^j)-2(\phi_n)^2\phi^i\phi^j\right],\label{S_def}\\
{N_l}^{ijk}&=&e_1\epsilon^{ijklm}\phi_m+e_1e_2\epsilon^{ijklmn}\phi_m\phi^n\phi_l+3e_1^2\delta^{ijk}_{lmn}\phi_m\phi^n,\label{Nijkl_def}\\
N^i&=&-e_1^2\phi_i-e_1e_2(\phi_j)^2\phi^i
\end{eqnarray}
with
\begin{equation}
\delta^{ijk}_{lnm}=\delta^{[i}_l\delta^j_n\delta^{k]}_m\, .
\end{equation}
We emphasize that raising and lowering of $SU(5)$ indices $i,j,k,\ldots$ correspond to taking complex conjugate. The field strengths $G^+_{\mu\nu ij}$ are obtained from $F^+_{\mu\nu ij}$ by dressing with scalars
\begin{equation}
G^+_{\mu\nu ij}=S^{ij,kl}F^+_{\mu\nu kl}\, .
\end{equation}
Finally, the tensor ${C_{ij}}^{kl}$ is given by
\begin{equation}
{C_{ij}}^{kl}=\frac{1}{e_1}\delta^{kl}_{ij}-2\frac{e_2}{e_1}\delta^{[k}_{[i}\phi^{l]}_{j]}
\end{equation}
with $\delta^{ij}_{kl}=\delta^{[i}_k\delta^{j]}_l$.
\\
\indent At this point, we note that we have corrected a typo in the sign of the second term in \eqref{Nijkl_def} as given in \cite{de_Wit_N5}. This sign change is also required by the supersymmetric Ward identity according to which the scalar potential can be written in term of the fermion-shift matrices as
\begin{equation}
V=-\frac{1}{5}g^2\left(6S^{ij}S_{ij}-\frac{1}{3}{N_l}^{ijk}{N^l}_{ijk}-2N^iN_i\right).
\end{equation}
As already mentioned, this scalar potential has been studied in \cite{Warner_Potential}. There are two $AdS_4$ vacua, one with $N=5$ supersymmetry and the other one with completely broken supersymmetry. These two vacua are given respectively by
\begin{equation}
\phi^i=0,\qquad V_0=-6g^2,\qquad L=\frac{1}{\sqrt{2}g}
\end{equation}
and
\begin{eqnarray}
& &\phi^i=0,\quad i=1,2,3,\qquad \phi^4=-i\phi^5=\sqrt{\frac{2}{5}},\nonumber \\
& &V_0=-14g^2,\qquad L=\frac{\sqrt{3}}{\sqrt{14}g}\, .
\end{eqnarray}
We have denoted the cosmological constant by $V_0$. The supersymmetric critical point preserves the full $SO(5)$ gauge symmetry while the non-supersymmetric one is only invariant under $SO(3)\subset SO(5)$. The $AdS_4$ radius $L$ is related to the cosmological constant by
\begin{equation}
L^2=-\frac{3}{V_0}\, .
\end{equation}
We have also taken $g>0$ for definiteness. At the supersymmetric $AdS_4$ vacuum, all scalars have masses $m^2L^2=-2$ corresponding to operators of dimensions $\Delta=1,2$ in the dual $N=5$ SCFT. These operators are given by scalar and fermion bilinears (mass terms), respectively. Although we will not further consider the non-supersymmetric vacuum, it is useful to emphasize that it is stable in the sense that all scalar masses are above the BF bound $m^2L^2=-\frac{9}{4}$. The full mass spectrum can be read off from the $SO(3)\times SO(3)$ $AdS_4$ critical point of the maximal $N=8$ gauged supergravity given in \cite{Warner_Fisch} keeping only $SO(3)$ singlet scalars.
\\
\indent In subsequent sections, we will look for various types of supersymmetric solutions which are asymptotic to the $N=5$ supersymmetric $AdS_4$ vacuum.

%%%%%%%%%%%%%%%%%%%%%%%%%%%%%%%%%%%%%%%%%%%%%%%%%%%%%%%%%%%%%%%%%%%%%%%%%%%%%%%%%%%%%%%%%%%%%%%%%%%%%%%%%%%%%%%%%%%%%%%%%%%%%%%%%%%%%%%%%
\section{Holographic RG flows}\label{RG_flow}
We begin with holographic RG flow solutions in the form of domain walls interpolating between the supersymmetric $AdS_4$ vacuum in the UV and singular geometries in the IR. The metric ansatz is given by
\begin{equation}
ds^2=e^{2A(r)}dx^2_{1,2}+dr^2
\end{equation}
with $dx^2_{1,2}$ being the flat metric on three-dimensional Minkowski space. Scalar fields will depend only on the radial coordinate $r$, and all the other fields will be set to zero. We will also use Majorana representation for gamma matrices in which all $\gamma^\mu$ are real, but $\gamma_5$ is purely immaginary.

\subsection{RG flows with $SO(4)$ symmetry}
We first consider solutions with $SO(4)$ symmetry. Among the five complex scalars, only one scalar is an $SO(4)\subset SO(5)$ singlet. We will choose this singlet to be $\phi^5$ and set
\begin{equation}
\phi^5=\phi=\varphi e^{i\zeta},\qquad \phi^i=0,\quad i=1,2,3,4
\end{equation}
with real scalars $\varphi$ and $\zeta$. In this case, the tensor $S^{ij}$ is real and given by
\begin{equation}
S^{ij}=\frac{g}{\sqrt{1-|\phi|^2}}\delta^{ij}=\frac{\mc{W}}{\sqrt{2}}\delta^{ij}
 \end{equation}
in which we have introduced the ``superpotential'' $\mc{W}$ for convenience. In general, the function $\mc{W}$ is complex and, in the present case of $N=5$ gauged supergravity, related to the eigenvalue $s$ of $S^{ij}$ corresponding to the unbroken supersymmetry by
\begin{equation}
 \mc{W}=\sqrt{2}s\, .
 \end{equation}
 The form of $S^{ij}$ being proportional to the identity matrix indicates that the entire flow will preserve $N=5$ supersymmetry with all $\epsilon^i$ non-vanishing.
 \\
 \indent Since there is an $r$-dependence in both the warped factor $A$ and scalar $\phi$, the gamma matrix $\gamma_{\hat{r}}$ will appear in the resulting BPS conditions. We then impose the following projector
\begin{equation}
\gamma_{\hat{r}}\epsilon_i=e^{i\Lambda}\epsilon^i\label{gamma_r_Pro}
 \end{equation}
with $\Lambda$ being a real function of $r$. Note that this projector relates the two chiralities of $\epsilon^i$, so the flow solutions will preserve only half of the original supersymmetry or ten supercharges. It is also possible that $\epsilon^i$ in different representations of the residual symmetry can have different phases.
 \\
 \indent With all these, the conditions $\delta\psi^i_\mu=0$ for $\mu=0,1,2$ give
 \begin{equation}
 e^{i\Lambda}A'+\mc{W}=0\, .
  \end{equation}
 Throughout this paper, we use $'$ to denote $r$-derivatives. This equation leads to
 \begin{equation}
 A'=\pm W= \pm |\mc{W}|\qquad \textrm{and}\qquad e^{i\Lambda}=\mp \frac{\mc{W}}{W}\, .
 \end{equation}
 In what follow, we will make a definite sign choice by choosing the upper signs in order to bring the supersymmetric $AdS_4$ critical point at $r\rightarrow \infty$.
 \\
 \indent The variations $\delta \chi^{ijk}$ and $\delta \chi$ lead to two equations of the form 
 \begin{equation}
 e^{-i\Lambda}\phi'=\sqrt{2}g\phi\sqrt{1-|\phi|^2}\qquad \textrm{and}\qquad  e^{-i\Lambda}{\phi^*}'=\sqrt{2}g\phi^*\sqrt{1-|\phi|^2}\, .
  \end{equation}
In the present case, $\mc{W}$ is real leading to $e^{i\Lambda}=\mp 1$, and we simply have the BPS equations
\begin{eqnarray}
A'&=&\frac{\sqrt{2}g}{\sqrt{1-|\phi|^2}},\\
\phi'&=&(1-|\phi|^2)^2\frac{\pd W}{\pd \phi^*}=-\sqrt{2}g\phi \sqrt{1-|\phi|^2}\, .
\end{eqnarray}
The scalar potential can also be written in term of $W$ as
\begin{equation}
V=4(1-|\phi|^2)^2\frac{\pd W}{\pd\phi}\frac{\pd W}{\pd \phi^*}-3W^2=-g^2\left(2+\frac{4}{1-|\phi|^2}\right).\label{SO4_V}
\end{equation}
\indent In terms of the real scalars $\varphi$ and $\zeta$, we simply have
\begin{equation}
A'=\frac{\sqrt{2}g}{\sqrt{1-\varphi^2}},\qquad \varphi'=-\sqrt{2}g\varphi\sqrt{1-\varphi^2},\qquad \zeta'=0\, .
\end{equation}
It is straightforward to verify that these equations are compatible with the corresponding field equations. The last equation together with the fact that $\zeta$ does not appear in any equations imply that $\zeta$ can be any constant without affecting the resulting solutions. We will choose $\zeta=0$ for definiteness. We also note that the condition $\delta \psi^i_{\hat{r}}=0$ gives the usual form of the Killing spinors for domain walls
\begin{equation}
 \epsilon^i=e^{\frac{A}{2}}\epsilon^i_{(0)}
\end{equation}
for spinors $\epsilon^i_{(0)}$ satisfying the projector \eqref{gamma_r_Pro}.
\\
\indent The flow solution can be readily obtained
\begin{eqnarray}
A&=&\frac{1}{2}\ln (1-\varphi^2)-\ln \varphi,\\
\varphi&=&\frac{2e^{\sqrt{2}gr-C}}{e^{2(\sqrt{2}gr-C)}+1}
\end{eqnarray}
with $C$ being an integration constant. It should be noted that $C$ can be set to zero by shifting the coordinate $r$. We have also neglected an irrelevant additive integration constant in $A$ since this can be removed by rescaling coordinates on $dx^2_{1,2}$. We now consider asymptotic behaviors of the solution. As $r\rightarrow \infty$, we find, recall that $L=\frac{1}{\sqrt{2}g}$,
\begin{equation}
\varphi\sim e^{-\sqrt{2}gr}\sim e^{-\frac{r}{L}}\qquad \textrm{and}\qquad A\sim \sqrt{2}gr\sim \frac{r}{L}\, .
\end{equation}
This is the $N=5$ supersymmtric $AdS_4$ configuration.
\\
\indent Furthermore, there is a singularity as $r\rightarrow \frac{C}{\sqrt{2}g}$ at which the solution becomes
\begin{equation}
\varphi\sim 1-\frac{1}{2}(\sqrt{2}gr-C)^2\qquad \textrm{and}\qquad A\sim \ln(\sqrt{2}gr-C).
\end{equation}
Near the singularity, we find that $\varphi\rightarrow 1$, $A\rightarrow -\infty$ and
\begin{equation}
V\sim -\frac{4g^2}{1-\varphi^2}\sim -\frac{4g^2}{(\sqrt{2}gr-C)^2}\rightarrow -\infty.
\end{equation}
 According to the criterion given in \cite{Gubser_singularity}, the singularity is then physically acceptable. Therefore, the above solution describes an RG flow from the $N=5$ SCFT in the UV to an $N=5$ non-conformal field theory in the IR. The flow breaks conformal symmetry but preserves the full $N=5$ Poincare supersymmetry in three dimensions. We identify this flow with the mass deformation pointed out in \cite{N5_6_3D_SCFT} in which the explicit form of relevant mass terms have also been given. The deformation preserves $N=5$ supersymmetry but breaks the $SO(5)$ R-symmetry to an $SO(4)$ subgroup in agreement with the supergravity result obtained here.

\subsection{RG flows with $SO(3)$ symmetry}
\begin{figure}
  \centering
  \begin{subfigure}[b]{0.45\linewidth}
    \includegraphics[width=\linewidth]{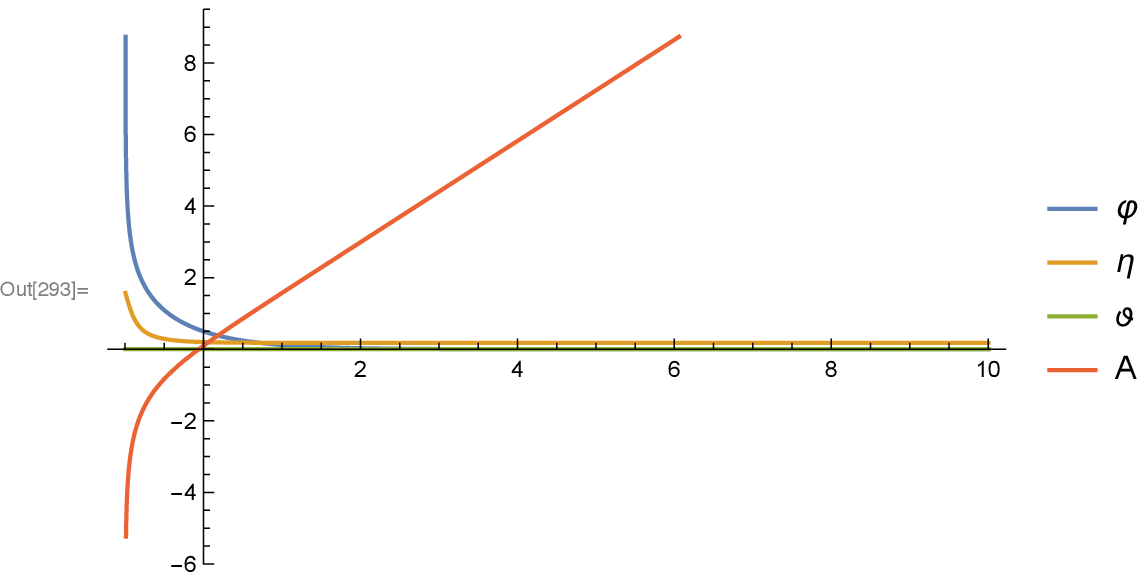}
  \caption{Flow I: An RG flow solution with $\vartheta=0$ along the flow.}
  \end{subfigure}
  \begin{subfigure}[b]{0.45\linewidth}
    \includegraphics[width=\linewidth]{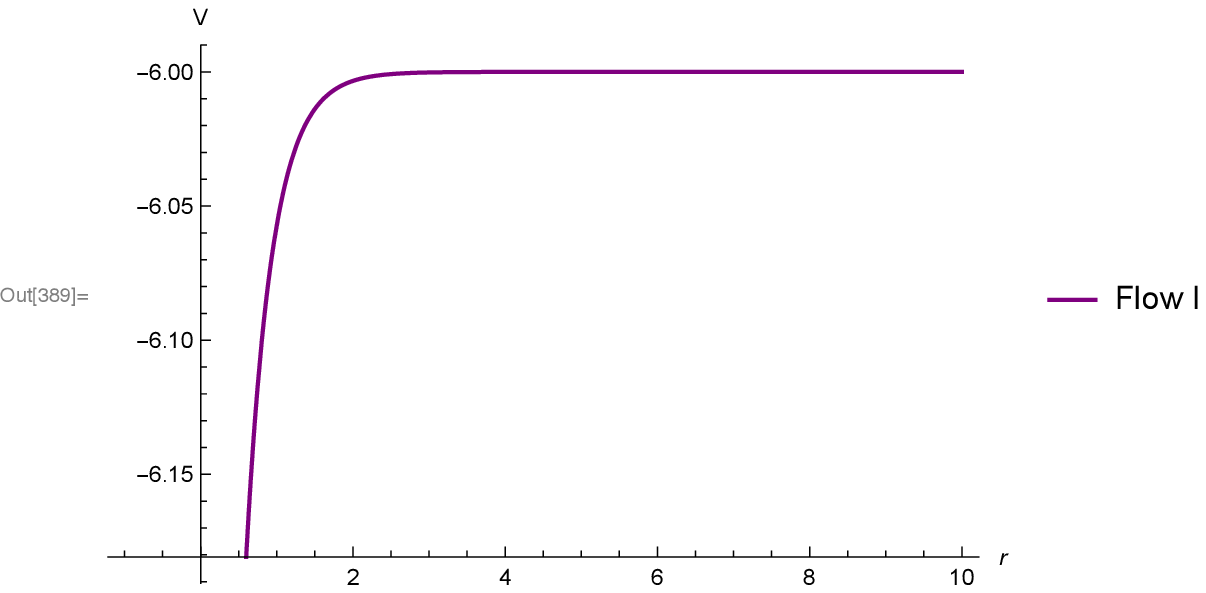}
  \caption{The behavior of the scalar potential along flow I.}
  \end{subfigure}
  \caption{An RG flow from the $N=5$ $AdS_4$ critical point as $r\rightarrow\infty$ to a non-conformal field theory in the IR with $\vartheta=0$.}
  \label{fig1}
\end{figure}

\begin{figure}
  \centering
  \begin{subfigure}[b]{0.45\linewidth}
    \includegraphics[width=\linewidth]{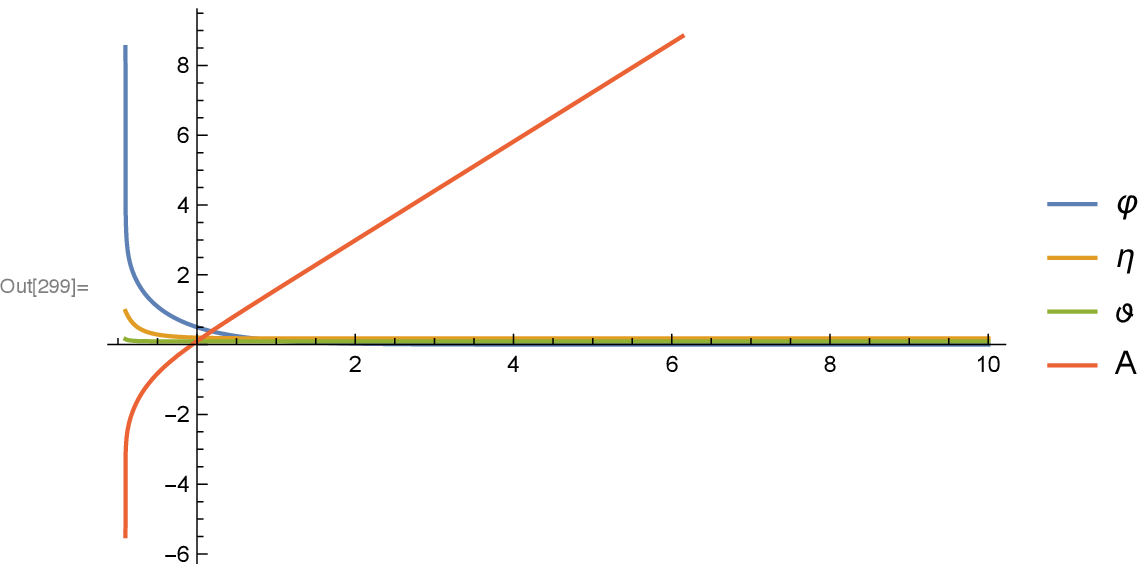}
  \caption{Flow II: An RG flow solution with $\vartheta\neq 0$ and $\varphi\rightarrow \infty$ in the IR.}
  \end{subfigure}
  \begin{subfigure}[b]{0.45\linewidth}
    \includegraphics[width=\linewidth]{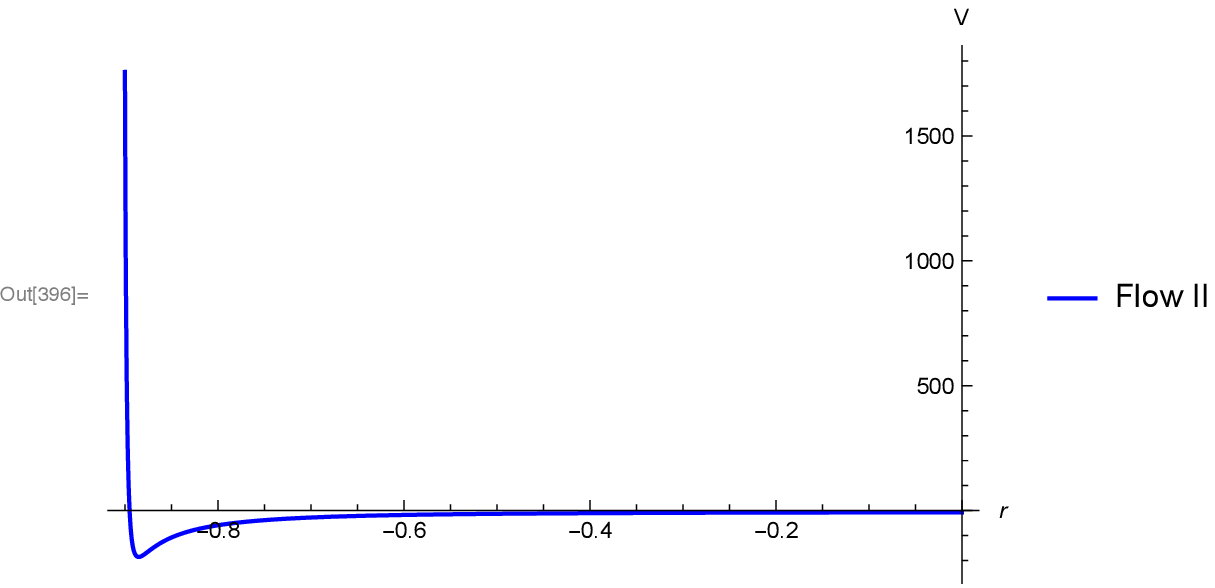}
  \caption{The behavior of the scalar potential along flow II.}
  \end{subfigure}
  \caption{An RG flow from the $N=5$ $AdS_4$ critical point as $r\rightarrow\infty$ to a non-conformal field theory in the IR with $\vartheta\neq 0$ and $\varphi\rightarrow \infty$ in the IR.}
  \label{fig2}
\end{figure}

\begin{figure}
  \centering
  \begin{subfigure}[b]{0.45\linewidth}
    \includegraphics[width=\linewidth]{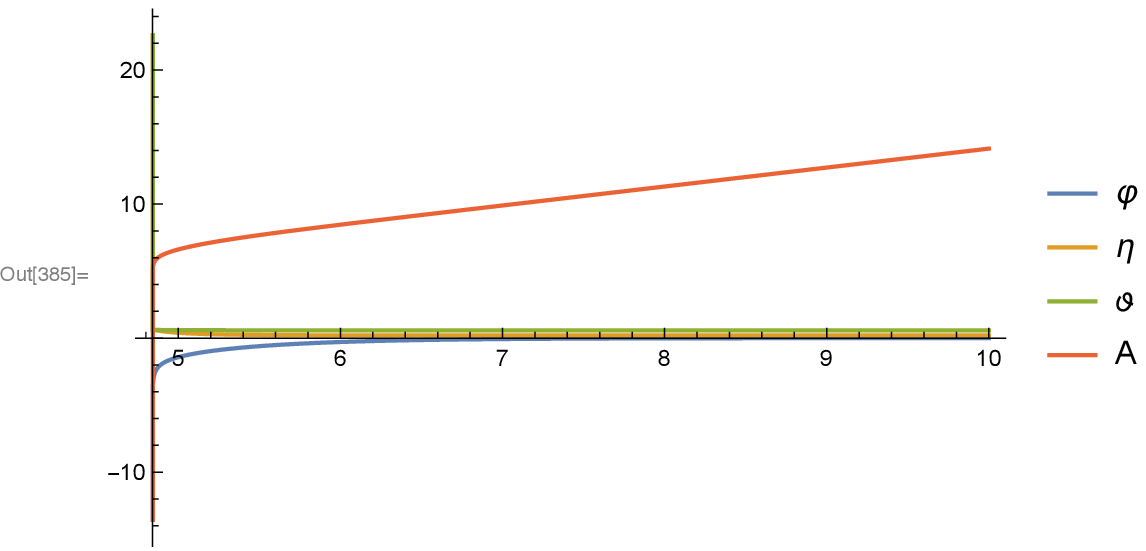}
  \caption{Flow III: An RG flow solution with $\vartheta\neq 0$ and $\varphi\rightarrow -\infty$ in the IR.}
  \end{subfigure}
  \begin{subfigure}[b]{0.45\linewidth}
    \includegraphics[width=\linewidth]{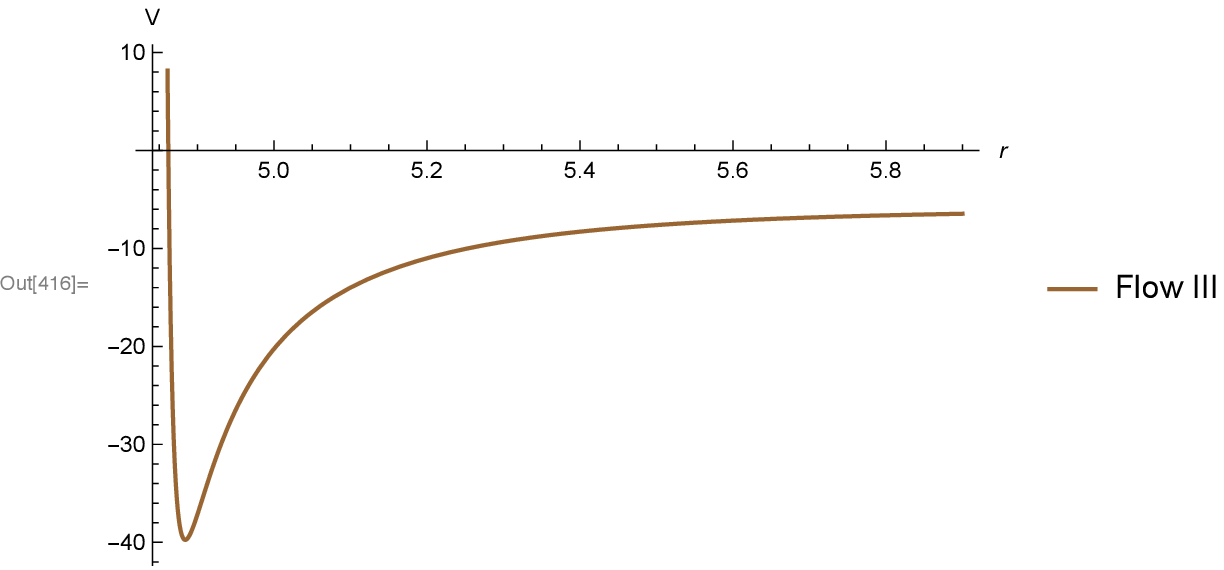}
  \caption{The behavior of the scalar potential along flow III.}
  \end{subfigure}
  \caption{An RG flow from the $N=5$ $AdS_4$ critical point as $r\rightarrow\infty$ to a non-conformal field theory in the IR with $\vartheta\neq 0$ and $\varphi\rightarrow -\infty$ in the IR.}
  \label{fig3}
\end{figure}

We now repeat the analysis for a smaller residual symmetry $SO(3)\subset SO(5)$. There are two scalars which are $SO(3)$ singlets. We will choose these scalars to be $\phi^4$ and $\phi^5$. It is more convenient to use the scalars in the form of
\begin{equation}
\phi^4=\tanh\varphi \cos \vartheta e^{i\zeta_1}\qquad \textrm{and}\qquad \phi^5=\tanh\varphi \sin \vartheta e^{i\zeta_2}\, .
\end{equation}
The variations $\delta \chi^{ijk}=0$ along $\epsilon^i$, $i=1,2,3$, lead to the condition
\begin{equation}
\sin (\zeta_1-\zeta_2)=0\label{SO3_Con}
\end{equation}
which implies
\begin{equation}
\zeta_1=\zeta_2+n\pi
\end{equation}
for an interger $n$. With this, the matrix $S^{ij}$ is real as in the $SO(4)$ case and given by
\begin{equation}
S^{ij}=g\cosh\varphi \delta^{ij}
\end{equation}
which leads to the superpotential
\begin{equation}
\mc{W}=\sqrt{2}g\cosh \varphi\, .
 \end{equation}
The resulting BPS equations read
\begin{eqnarray}
A'&=&\sqrt{2}g\cosh\varphi,\qquad \varphi'=-\sqrt{2}g\sinh\varphi,\nonumber \\
\vartheta'&=&0,\qquad \zeta_1'=\zeta_2'=0\, .
\end{eqnarray}
The solutions for $A$ and $\varphi$ are the same as in the $SO(4)$ case, up to some field redefinition, with constant values of $\vartheta$ and $\zeta_{1,2}$ that can be chosen to be zero. Therefore, if we keep supersymmetry corresponding to $\epsilon^{1,2,3}$ unbroken, we necessarily find $N=5$ supersymmetric solutions with $SO(4)$ symmetry.
\\
\indent We now consider another possibility obtained by setting $\epsilon^i=0$ for $i=1,2,3$. The condition \eqref{SO3_Con} is then not needed. The remaining two eigenvalues of $S^{ij}$ are given by
\begin{eqnarray}
        \mathcal{W}_{\pm}&=&\frac{g}{4\sqrt{2}} \left[ 2\left(3+\cos{2\eta}\right)\cosh{\varphi} + \left( 3+\cosh{2\varphi} \right) \sin^2{\eta}\phantom{\frac{\varphi}{2}}\right. \nonumber \\
      &  &\left.-8\sinh^4{\frac{\varphi}{2}}\left( \cos{4\vartheta}\sin^2{\eta}\pm i \Gamma \right) \right]
\end{eqnarray}
with
\begin{equation}
    \Gamma = \sin{\eta} \sin{2\vartheta}\sqrt{3+\cos{2\eta}+2\cos{4\vartheta}\sin^2{\eta}}\, .
\end{equation}
The corresponding Killing spinors are given by
\begin{equation}
    \epsilon_{\pm}= \epsilon_{5} \pm \left(\frac{\sin{2\eta}\sin^2{2\vartheta}-\Gamma}{\sin{\eta}\sin{4\vartheta}}\right) \epsilon_{4}\, .
\end{equation}
We have redefined the scalars $\zeta_{1,2}$ by writting $\zeta_1=\psi$ and $\zeta_2=\psi-\eta$. 
\\
\indent Since $\Gamma$ is real, both $\mc{W}_+$ and $\mc{W}_-$ give the same real superpotential $W=|\mc{W}_\pm|$ in term of which the scalar potential can be written as
\begin{eqnarray}
        V &=&\left(\frac{\partial W}{\partial \varphi}\right)^2 + \frac{1}{\sinh^2{\varphi}}\left(\frac{\partial W}{\partial \vartheta}\right)^2 + \frac{4}{\sin^2{2\vartheta}\sinh^2{\varphi}}\left(\frac{\partial W}{\partial\eta}\right)^2-3W^2\nonumber \\
       &=& \frac{1}{2}g^2 \left( -8-4\cosh{2\varphi} +\sin^2{\eta}\sin^2{2\vartheta}\sinh^4{\varphi} \right).\label{SO3_V}
\end{eqnarray}
For general non-vanishing $\epsilon_\pm$, the solutions will preserve $N=2$ supersymmetry.
\\
\indent By imposing the following projectors
\begin{equation}
\gamma_{\hat{r}}\epsilon_\pm=e^{\pm i \Lambda}\epsilon^{\pm},
\end{equation}
we obtain the BPS equations
\begin{eqnarray}
    A' &=& W
    = \frac{g}{8}\left[ 67+\cosh{4\varphi}-16\cos{2\eta} \left(3+4\cosh{\varphi}\right)\sinh^4{\frac{\varphi}{2}}\right.\nonumber
    \\
  &  &\left.+ \cosh{2\varphi} \left(60-16\cos{2 \eta}\sinh^4{\frac{\varphi}{2}} \right)-16 \cos{4\vartheta}\sin^2{\eta}\sinh^4{\varphi} \right]^{1/2},\\
   \varphi' &=& - \frac{\partial W}{\partial \varphi}
        =\frac{g^2}{32 W} \left[ 8\cosh{\varphi}\left( \cos{2\eta}+2\cos{4\vartheta}\sin^2{\eta}\right)\sinh^3{\varphi}\right.
       \nonumber  \\
       & &\left.-30\sinh{2\varphi}-\sinh{4\varphi}  \right],\label{varphi_eq}\\
    \vartheta' &=& -\frac{1}{\sinh^2{\varphi}}\frac{\partial W}{\partial \vartheta}
        = -\frac{g^2}{2 W}\sin^2{\eta}\sin{4\vartheta}\sinh^2{\varphi},\label{vartheta_eq}\\
    \eta' &=& -\frac{4}{\sin^2{2\vartheta}\sinh^2{\varphi}}\frac{\partial W}{\partial \eta} = -\frac{g^2}{W} \sin{2\eta}\sinh^2{\varphi},\label{eta_eq}\\
    \psi'&=&0\, .
\end{eqnarray}
As in the previous case, we can set $\psi=0$ for convenience.
\\
\indent We are not able to completely solve these equations in an analytic form. However, by combining equations \eqref{vartheta_eq} and \eqref{eta_eq}, we obtain
\begin{equation}
\cot 2\vartheta=C_1\cos\eta
\end{equation}
with an integration constant $C_1$. Similarly, combining equations \eqref{varphi_eq} and \eqref{eta_eq} gives
\begin{eqnarray}
2\sqrt{2}\textrm{sech}^2\varphi&=&32(1+C_1^2)C_2\sqrt{(1+\cos2\eta)(2+C_1^2+C_1^2\cos2\eta)}\nonumber  \\
& &-3-4C_1^2\cos\eta-\cos2\eta 
\end{eqnarray}
with another integration constant $C_2$. In addition, for a particular value of $C_2=0$, we find the solution for $A$ as follows
\begin{equation}
 A=2\ln(2+C_1^2+C_1^2\cos 2\eta)-2\ln(3+2C_1^2\cos2\eta).
\end{equation}
\indent The complete solution can be obtained numerically. Examples of these solutions are given in figures \ref{fig1}, \ref{fig2} and \ref{fig3} with $g=1$. For convenience, we will call the solutions shown in these figures flow I, flow II and flow III, respectively. For flow I, $\vartheta$ vanishes along the entire flow. This is nothing but the $SO(4)$ symmetric flow solution given in the previous section. From the behavior of the scalar potential, it is clearly seen that the IR singularity is physical in agreement with the previous result.
\\
\indent For flow II and flow III with $\vartheta\neq 0$ along the flows, we find that near the singularities, $\varphi\rightarrow \infty$ and $\varphi\rightarrow -\infty$, respectively. Both of these flows are unphysical by the criterion of \cite{Gubser_singularity} since the scalar potential goes to infinity near the singularities. This behavior can also be seen from the potential given in \eqref{SO3_V}. For $\varphi\rightarrow \pm \infty$, we find that
\begin{equation}
V\sim \sin^2\eta\sin^2\vartheta e^{\pm 4\phi}\rightarrow \infty
\end{equation}
unless $\sin\vartheta=0$ or $\sin\eta=0$ which give the $SO(4)$ symmetric solution.

\indent We end this section by giving a solution for particular values of $\eta=\frac{\pi}{2}$ and $\vartheta=\frac{\pi}{4}$. The resulting BPS equations become
\begin{equation}
A'=\frac{g}{2\sqrt{2}}(2+\cosh 2\varphi),\qquad \varphi'=-\frac{g}{\sqrt{2}}\sinh\varphi,\qquad \eta'=\vartheta'=\psi'=0\, .
\end{equation}
The solution can be readily obtained
\begin{eqnarray}
\tanh \varphi&=&e^{-\sqrt{2}g(r-r_0)},\\
A&=&\frac{1}{2}\ln \cosh\varphi-\ln \sinh\varphi\, .
 \end{eqnarray}
Although this is very similar to the $SO(4)$ symmetric solution, it should be noted that this solution only preserves $N=2$ supersymmetry in three dimensions and breaks $SO(5)$ to $SO(3)$. We recall that when $\eta=0$, the unbroken supersymmetry is enhanced to $N=5$ as previously mentioned. However, the values $\eta=\frac{\pi}{2}$ and $\vartheta=\frac{\pi}{4}$ lead to $V\rightarrow \infty$ near the singularities as $\varphi\rightarrow \pm\infty$. It would be interesting to find an uplift of this solution to M-theory using the $S^7$ truncation to the maximal $N=8$ gauged supergravity and check whether the singularities are acceptable. If this is the case, identifying the analogue of non-vanishing $\eta$ and $\vartheta$ in the dual $N=5$ SCFT that breaks $N=5$ supersymmetry to $N=2$ also deserves further study.

\subsection{RG flows with $SO(2)$ symmetry}
We finally consider the smallest possible residual symmetry $SO(2)\subset SO(5)$. There are three singlet scalars which are chosen to be
\begin{equation}
\phi^i=\varphi_i e^{i\zeta_i},\qquad i=3,4,5\, .
\end{equation}
\indent Similar to the $SO(3)$ case, the BPS conditions along $\epsilon^1$ and $\epsilon^2$ lead to the conditions
\begin{equation}
\zeta_i-\zeta_j=n\pi
\end{equation}
for any interger $n$ and $i\neq j$. If we impose this condition, the solutions will preserve $N=5$ supersymmetry subject to the $\gamma_{\hat{r}}$ projector. On the other hand, we can, as in the previous case, set $\epsilon^{1,2}=0$ and look for solutions with at most $N=3$ supersymmetry. The BPS equations in this case are much more complicated than the $SO(3)$ case due to more scalars involved. Therefore, we will only consider the simpler situation of setting $\zeta_i=\zeta$ for $i=3,4,5$.
\\
\indent By the same analysis as in the previous cases together with $\zeta_1=\zeta_2=\zeta_3=\zeta$, we find that $\zeta'=0$. We can again set $\zeta=0$ and find the following BPS equations
\begin{eqnarray}
A'&=&\frac{\sqrt{2}g}{\sqrt{1-\varphi_3^2-\varphi_4^2-\varphi_5^2}},\\
\varphi_i'&=&-\sqrt{2}g\varphi_i\sqrt{1-\varphi_3^2-\varphi_4^2-\varphi_5^2},\qquad i=3,4,5\, .
\end{eqnarray}
These equations imply $\varphi_3=\alpha \varphi_5$ and $\varphi_4=\beta\varphi_5$ for constants $\alpha$ and $\beta$. Using this fact and rewritting $\phi^5=\varphi$, we end up with only two equations for $A'$ and $\varphi'$. The corresponding solution is given by
\begin{eqnarray}
A&=&\sqrt{2}gr+\ln\left[1-(1+\alpha^2+\beta^2)e^{2(C-\sqrt{2}gr)}\right],\\
\varphi&=&\frac{2e^{C-\sqrt{2}gr}}{1+(1+\alpha^2+\beta^2)e^{2(C-\sqrt{2}gr)}}\, .
 \end{eqnarray}
The singularity when $2\sqrt{2}gr\sim \ln[e^{2C}(1+\alpha^2+\beta^2)]$ at which $A\rightarrow -\infty$ and
\begin{equation}
\varphi\sim \frac{1}{\sqrt{1+\alpha^2+\beta^2}}
\end{equation}
is also physically acceptable since
\begin{equation}
V\sim -\frac{4g^2}{1-(1+\alpha^2+\beta^2)\varphi^2}\rightarrow -\infty
\end{equation}
near the singularity.
\\
\indent If we redefine the scalar to $\tilde{\varphi}=\sqrt{1+\alpha^2+\beta^2}\varphi$, we find exactly the same $SO(4)$ symmetric solution given previously. Therefore, it appears that the only physical RG flow within the framework of $N=5$ gauged supergravity is the $SO(4)$ symmetric one preserving $N=5$ supersymmetry. Note also that from $SO(3)$ and $SO(2)$ symmetric solutions, we see that the solutions with only real scalars non-vanishing reduce to the $SO(4)$ symmetric solution preserving half of the original supersymmetry. We will show below that this result is indeed valid in general.     

\subsection{Comment on general supersymmetric domain wall solutions}
Since there are only five complex scalars in $N=5$ gauged supergravity, we can generalize the results obtained in the previous cases to the full $SU(5,1)/U(5)$ scalar coset. We first consider solutions with a residual symmetry $SO(n)$ for $1<n<5$. For $n=5$, no scalars can be turned on because there is no $SO(5)$ singlet among the five scalars. 
\\
\indent To proceed further, we recall that, with only scalar fields non-vanishing, the conditions $\delta \chi^{ijk}=0$ from \eqref{d_chi} can be written as
\begin{equation}\label{chide}
    \delta \chi^{ijk}= -\epsilon^{ijklm}\gamma^{\mu}  P_{\mu m}\epsilon_l - 2g {N_{l}}^{ijk}\epsilon^{l}=0 
\end{equation}
with
\begin{equation} \label{A2}
    {N_{l}}^{ijk}=e_1 \epsilon^{ijklm}\phi_m + e_1 e_2 \epsilon^{ijkmn}\phi_m \phi^n \phi_l + 3 e_1^2 \delta_{lmn}^{ijk}\phi_m\phi^n\, .
\end{equation}
\indent It turns out that some of these conditions do not involve derivatives of scalars from $P^i_\mu$. In particular, this can happen when indices $l$ and $i$ are equal among other possibilities. We then write $\phi^m = \varphi_m e^{i \zeta_m}$ and consider the conditions $\delta\chi^{ijk}=0$, for $l=i$, which reduce to
\begin{equation}
e_1 e_2 \epsilon^{ljkmn}\varphi_m \varphi_n \varphi_l e^{i (-\zeta_m+\zeta_n-\zeta_l)}+ 3 e_1^2 \delta_{lmn}^{ljk}\varphi_m\varphi_n e^{i(-\zeta_m+\zeta_n)}=0
\end{equation}
without summing over $l$. By antisymmetrizing the products of $\varphi_m$'s, we arrive at the result
\begin{equation}
e_1 e_2\epsilon^{ljkmn} \varphi_l  \varphi_n \varphi_m e^{-i \zeta_l} \sin{(\zeta_n - \zeta_m)} + 6e_1^2 \varphi_j \varphi_k \sin{(\zeta_j -\zeta_k)}=0\, .
\end{equation}
Since the two terms on the left hand side are independent of each other, this condition implies  
\begin{equation} \label{sinpsi}
    \sin(\zeta_{i} - \zeta_{j}) =0
\end{equation}
which gives the previously obtained result $\zeta_i=\zeta_j+n\pi $.
\\
\indent By splitting indices $i,j,\ldots=1,2,\ldots, 5$ into $\hat{i},\hat{j},\ldots =1,2,\ldots,n$ and $\tilde{i},\tilde{j},\ldots =n+1,\ldots, 5$ with scalars $\varphi^{\tilde{i}}$ and $\phi^{\hat{i}}$ being respectively singlets and non-singlets of $SO(n)$, we can summarize possible cases as follow.
\begin{itemize}
    \item For $n=4$, there is only one $SO(4)$ singlet scalar, and in this case ${N_{l}}^{ljk}$ automatically vanish.
    \item For $1<n<4$, there are $5-n$ singlet scalars denoted by $\phi^{\tilde{i}}$. The relevant non-vanishing components of ${N_l}^{ijk}$ are ${N_{\hat{l}}}^{\hat{l}\tilde{j}\tilde{k}}$ which lead to the conditions $\sin(\zeta_{\tilde{i}}-\zeta_{\tilde{j}})=0$. Accordingly, we need to set $\zeta_{\tilde{i}}=\zeta_{\tilde{j}}+m\pi$ or $\epsilon_{\hat{i}}=0$. In the former case, all the phases are equivalent up to an additive constant $m\pi$ and lead to the tensor $S^{ij}$ proportional to the identity matrix, possibly after a diagonalization. The latter case gives $N=5-n$ supersymmetric solutions with the corresponding Killing spinors $\epsilon^{\tilde{i}}$.
\end{itemize}
In particular, this result implies that domain wall solutions with all five scalars non-vanishing are possible only when all the complex phases of the scalars are equal up to an additive constant $m\pi$. In addition, for scalar fields of the form
\begin{equation} 
\phi^i=(\varphi_1,\varphi_2,\varphi_3,\varphi_4,\varphi_5)e^{i\zeta}
\end{equation}
with $m=0$ for convenience, we can verify from the definition \eqref{S_def} that the $S^{ij}$ tensor is real and independent of $\zeta$. This leads to the BPS equation $\zeta'=0$ according to which $\zeta$ can be set to zero. 
\\
\indent Furthermore, by using the parametrization of the form
\begin{eqnarray}
\varphi_1&=&\varphi \cos\xi_1,\qquad \varphi_2=\varphi\sin\xi_1\cos\xi_2,\qquad \varphi_3=\varphi \sin\xi_1\sin \xi_2\cos\xi_3,\nonumber \\
\varphi_4&=&\varphi\sin\xi_1\sin\xi_2\sin\xi_3\cos\xi_4,\qquad  \varphi_5=\varphi\sin\xi_1\sin\xi_2\sin\xi_3\sin\xi_4,
\end{eqnarray}
we readily find
\begin{equation}
S^{ij}=\frac{g}{\sqrt{1-\varphi^2}}
\end{equation}
with the scalar potential
\begin{equation}
V=-\frac{2g^2(3-\varphi^2)}{1-\varphi^2}\, .
\end{equation}
Therefore, the resulting BPS equations will give $\xi_i'=0$ for all $i=1,2,3,4$. Since only $\varphi$ depends on the radial coordinate $r$, the solution effectively reduces to that of the $SO(4)$ case. We can then conclude that the most general half-supersymmetric domain wall solutions of $N=5$ gauged supergravity can only involve non-vanishing real scalars with $SO(4)$ symmetry. However, we note here that this conclusion is valid only for half-BPS solutions. More general flow solutions with less supersymmetry are possible but these solutions necessarily involve non-vanishing pseudoscalars. 

%%%%%%%%%%%%%%%%%%%%%%%%%%%%%%%%%%%%%%%%%%%%%%%%%%%%%%%%%%%%%%%%%%%%%%%%%%%%%%%%%%%%%%%%%%%%%%%%%%%%%%%%%%%%%%%%%%%%%%%%%%%%%%%%%%%%%%%%%
\section{Supersymmetric Janus solutions}\label{Janus_N5}
We now move to supersymmetric Janus solutions obtained from an $AdS_3$-sliced domain wall ansatz
\begin{equation}
ds^2=e^{2A}(e^{\frac{2\xi}{\ell}}dx^2_{1,1}+d\xi^2)+dr^2\, .
\end{equation}
The analysis is essentially the same as that given in \cite{warner_Janus}, see also \cite{N3_Janus}. Therefore, in this paper, we will mainly review relevant results for deriving the BPS equations.
\\
\indent In this case, the BPS equations will get modified compared to the RG flow case due to the curvature of the three-dimensional slices. The conditions $\delta \psi^i_{\hat{\mu}}=0$ for $\hat{\mu}=0,1$ give
\begin{equation}
A'\gamma_{\hat{r}}\epsilon_i+\frac{1}{\ell}e^{-A}\gamma_{\hat{\xi}}\epsilon_{i}+\mc{W}\epsilon^i=0\label{dPsi_mu_eq}
\end{equation}
which leads to the following equation
\begin{equation}
A'^2=W^2-\frac{1}{\ell^2}e^{-2A}\label{dPsi_BPS_eq}
\end{equation}
with $W=|\mc{W}|$ as usual. 
\\
\indent We then impose the $\gamma_{\hat{\xi}}$ projection of the form
\begin{equation}
\gamma_{\hat{\xi}}\epsilon_i=i\kappa e^{i\Lambda}\epsilon^i\label{gamma_xi_pro}
\end{equation}
with $\kappa^2=1$. The constant $\kappa=\pm 1$ defines the chirality of the Killing spinors on the two-dimensional conformal defects described by the $AdS_3$-slices. Using the projector \eqref{gamma_xi_pro} in equation \eqref{dPsi_mu_eq} lead to the $\gamma_{\hat{r}}$ projector given in \eqref{gamma_r_Pro} with the phase factor
\begin{equation}
e^{i\Lambda}=-\frac{A'}{W}-\frac{i\kappa}{\ell}\frac{e^{-A}}{W}\label{real_W_phase} 
\end{equation}
for real $\mc{W}$ and
\begin{equation}
e^{i\Lambda}=-\frac{\mc{W}}{A'+\frac{i\kappa}{\ell}e^{-A}}\label{complex_W_phase}
\end{equation}
for complex $\mc{W}$. It should be noted that the terms involving the superpotential have opposite signs to those given in \cite{warner_Janus} and \cite{N3_Janus} due to the different definition of the superpotential in term of the eigenvalue of $S^{ij}$. 
\\
\indent Taking into account the conditions coming from $\delta \psi^i_{\hat{\xi}}=0$ and $\delta\psi^i_{\hat{r}}=0$, we can derive the explicit form of the Killing spinors, see more detail in \cite{warner_Janus}, as follows
\begin{equation}
\epsilon_i=e^{\frac{A}{2}+\frac{\xi}{2\ell}+i\frac{\Lambda}{2}}\varepsilon^{(0)}_i
\end{equation}
where the constant spinors $\varepsilon^{(0)}_{i}$ satisfy
\begin{equation}
\gamma_{\hat{r}}\varepsilon^{(0)}_{i}=\varepsilon^{(0)i}\qquad
\textrm{and}\qquad
\gamma_{\hat{\xi}}\varepsilon^{(0)}_{i}=i\kappa\varepsilon^{(0)i}\, .
\end{equation}
After using the $\gamma_{\hat{r}}$ projector in the variations $\delta\chi$ and $\delta\chi^{ijk}$, we obtain the full set of BPS equations. We emphasize again that different phases $e^{i\Lambda}$ for $\epsilon^i$ in different representations under a given residual symmetry are possible as also pointed out in \cite{warner_Janus}.    

\subsection{Janus solutions with $SO(4)$ symmetry}
We first give Janus solutions with $SO(4)$ symmetry under which $\epsilon^i$ transform as $\mathbf{4}+\mathbf{1}$. In order to obtain a consistent set of BPS equations, we need to impose the following projectors 
 \begin{eqnarray}
    \gamma_{\hat{r}}\epsilon_{\hat{i}}&=&e^{i\Lambda}\epsilon^{\hat{i}},\qquad \hat{i}=1,2,3,4,\qquad \gamma_{\hat{r}}\epsilon_{5}=e^{-i\Lambda}\epsilon^{5},\nonumber \\  
    \gamma_{\hat{\xi}}\epsilon_{\hat{i}}&=& i \kappa e^{i \Lambda}\epsilon^{\hat{i}},\quad \qquad  \gamma_{\hat{\xi}}\epsilon_{5}= -i \kappa e^{-i \Lambda}\epsilon^{5}\, . 
\end{eqnarray} 
Using the superpotential 
\begin{equation}
\mc{W}=\frac{\sqrt{2}g}{\sqrt{1-\varphi^2}}
\end{equation}
and the phase \eqref{real_W_phase}, we find the following BPS equations
\begin{eqnarray}
\varphi'&=&-\frac{2g^2\ell^2\varphi A'e^{2A}}{1+\ell^2A'^2e^{2A}},\label{JanusSO4_varphi_eq}\\
\zeta'&=&-\frac{2g^2\kappa\ell e^{A}}{1+\ell^2A'^2+e^{2A}},\label{JanusSO4_zeta_eq}\\
A'^2&=&\frac{e^{-2A}(2g^2\ell^2e^{2A}+\varphi^2-1)}{\ell^2(1-\varphi^2)}\, .\label{JanusSO4_A_eq}
\end{eqnarray}  
It should be noted that, in this case, the phase $\zeta$ is not constant along the flow. Furthermore, these equations reduce to those of the RG flow studied in the previous section in the limit $\ell\rightarrow \infty$ for which the $AdS_3$-slices become flat.
\\
\indent By combing equations \eqref{JanusSO4_varphi_eq} and \eqref{JanusSO4_A_eq}, we find
\begin{equation}
 \frac{dA}{d\varphi}=-\frac{1}{\varphi-\varphi^3}
\end{equation}   
which gives
\begin{equation}
A=\frac{1}{2}\ln(1-\varphi^2)-\ln\varphi\, .
\end{equation}
Using the solution for $A$ in equation \eqref{JanusSO4_varphi_eq}, we obtain the solution for $\varphi$ given by
\begin{eqnarray}
& &2\sqrt{2}gr=2\ln\varphi-\ln\left[2\sqrt{2}g\ell\sqrt{(1-\varphi^2)(2g^2\ell^2-\varphi^2)}-\varphi^2+2g^2\ell^2(2-\varphi^2)\right].\nonumber\\
& &
\end{eqnarray}
Finally, combining equations \eqref{JanusSO4_varphi_eq} and \eqref{JanusSO4_zeta_eq}, we find 
\begin{equation}
\zeta=\kappa \tan^{-1}\frac{\varphi}{\sqrt{2g^2\ell^2-\varphi^2}}+C
\end{equation}
for an integration constant $C$. We point out here that by redefining the scalar $\varphi$ as
\begin{equation}
\varphi=\tanh\tilde{\varphi},
\end{equation}
we obtain the same solution as given in \cite{warner_Janus} and \cite{N3_Janus} in $N=8$ and $N=3$ gauged supergravities. Therefore, this solution could be just the known solution of $N=8$ theory that survives the truncation to the $N=5$ theory. 
\\
\indent We end this section by giving a comment on the unbroken supersymmetry on the conformal defect. Since $\mc{W}$ is real, $e^{i\Lambda}$ and $e^{-i\Lambda}$ are related by a sign change in $\kappa$. This implies that $\epsilon^{\hat{i}}$ and $\epsilon^5$ are subject to the $\gamma_{\hat{\xi}}$ projector with opposite sign of $\kappa$, so the two-dimensional defect preserves $N=(4,1)$ or $N=(1,4)$ supersymmetry depending on the values of $\kappa=1$ or $\kappa=-1$, respectively. It is also useful to note the numbers of unbroken supersymmetries on the defect for solutions in $N=8$ and $N=3$ theories. These are given respectively by $N=(4,4)$ and $N=(2,1)$.

\subsection{Janus solutions with $SO(3)$ symmetry}
\begin{figure}
  \centering
  \begin{subfigure}[b]{0.3\linewidth}
    \includegraphics[width=\linewidth]{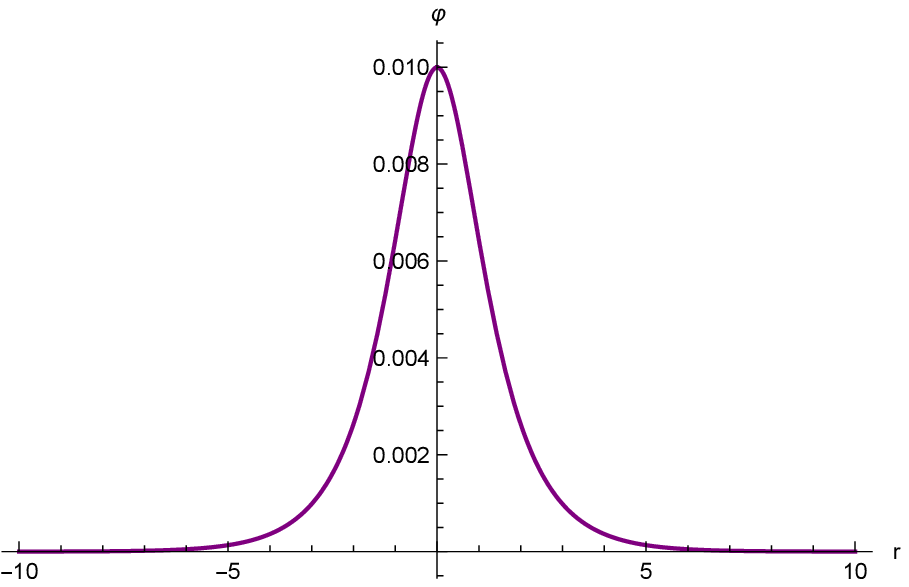}
  \caption{$\varphi$ solution.}
  \end{subfigure}
  \begin{subfigure}[b]{0.3\linewidth}
    \includegraphics[width=\linewidth]{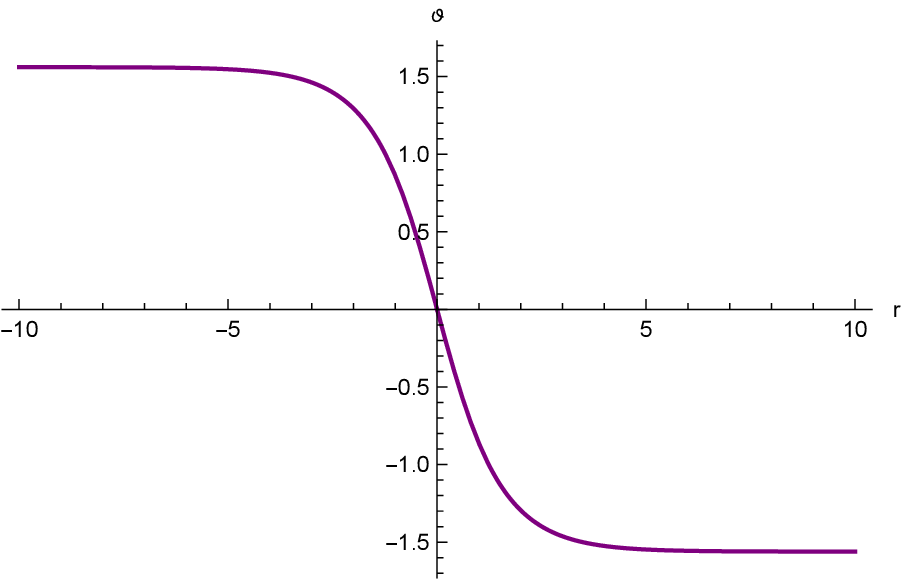}
  \caption{$\vartheta$ solution.}
  \end{subfigure}
  \begin{subfigure}[b]{0.3\linewidth}
    \includegraphics[width=\linewidth]{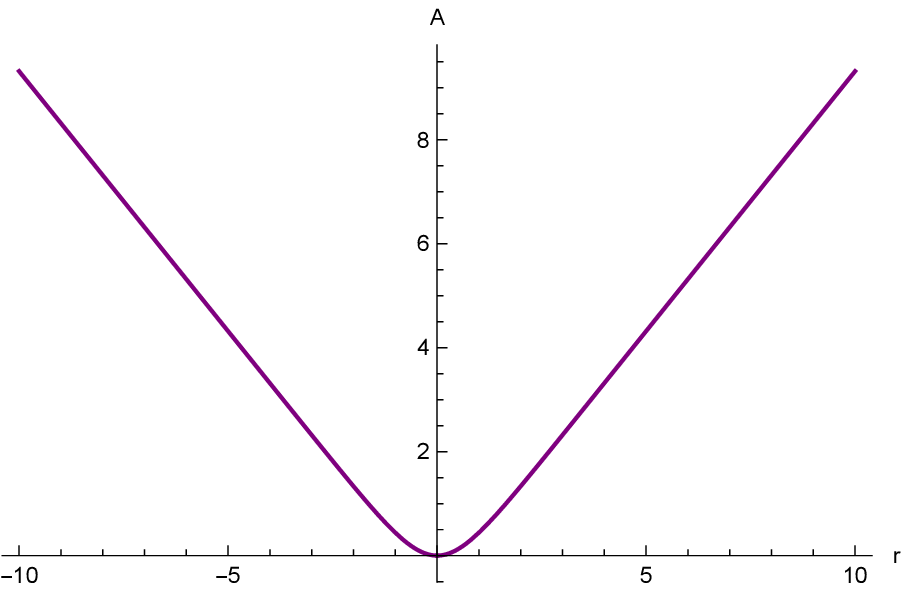}
  \caption{$A$ solution.}
  \end{subfigure}
  \caption{A Janus solution with $SO(3)$ symmetry and $N=(1,1)$ supersymmetry on the two-dimensional conformal defect within the $N=5$ SCFT.}
  \label{fig4}
\end{figure}

We now move to Janus solutions with $SO(3)$ symmetry with the corresponding singlet scalars given by
\begin{equation}
\phi^4=\tanh\varphi \cos\vartheta e^{i\zeta}\qquad\textrm{and}\qquad \phi^5=\tanh\varphi \sin \vartheta e^{i(\zeta-\eta)}\, . 
 \end{equation} 
For $\eta=0$, we find that the BPS conditions give $\vartheta'=0$, and the resulting BPS equations as well as the solution are the same as the $N=(4,1)$ Janus solution given in the previous section. 
\\
\indent As in the RG flow case, we look for different solutions with $\eta\neq 0$ by setting $\epsilon^{1,2,3}=0$. It turns out that the BPS equations for general values of $\eta$ are highly complicated. Therefore, we proceed by taking $\eta=\frac{\pi}{2}$ for simplicity. In this case, the superpotentials obtained from the eigenvalues of the $S^{ij}$ tensor are given by   
\begin{equation}
    \mathcal{W}_{\pm}= \sqrt{2} g \left( \cosh^4{\frac{\varphi}{2}}  - e^{\mp 4i \vartheta} \sinh^4{\frac{\varphi}{2}}\right)
\end{equation} 
corresponding to the Killing spinors $\epsilon_{\pm}=\epsilon_4\pm \epsilon_5$. By repeating the same analysis as in the previous case and imposing the following projectors on $\epsilon_{\pm}$
\begin{equation}
    \gamma^{\hat{r}}\epsilon_{\pm}=e^{\pm i \Lambda}\epsilon^{\pm}
    \qquad
    \textrm{and}
    \qquad
    \gamma^{\hat{\xi}}\epsilon_{\pm}=\pm i \kappa e^{\pm i \Lambda}\epsilon^{\pm},
\end{equation}
we  obtain the phase factor
\begin{equation} \label{two}
    e^{\pm i \Lambda}=-\frac{\mathcal{W}_{\pm}}{A' \pm i \frac{\kappa}{\ell}e^{-A}}\, .
\end{equation}
With this result, the conditions $\delta \chi^{ijk}=0$ and $\delta \chi =0$ lead to the BPS equations
\begin{eqnarray}
        \varphi' &=& -\left( \frac{A'}{W} \right) \frac{\partial W}{\partial \varphi} + \left(\frac{\kappa e^{-A}}{W \ell}\right)\frac{1}{\sinh{\varphi}}\frac{\partial W}{\partial \vartheta}\nonumber 
        \\
        &=& \frac{g^2}{16 W^2}\left[ 8\left(\frac{\kappa e^{-A}}{\ell}\right) \sin{4\vartheta}\sinh^3{\varphi}\right. \nonumber  
        \\
       & &\left.\phantom{\frac{e^{A}}{\ell}}+ A' \left(8\cos{4\vartheta}\cosh{\varphi}\sinh^3{\varphi}-14\sinh{2\varphi}-\sinh{4\varphi}\right) \right],\\
        \vartheta' &=& -\frac{1}{\sinh^2{\varphi}}\left(\frac{A'}{W}\right)\frac{\partial W}{\partial \vartheta} -\left(\frac{\kappa e^{-A}}{W \ell}\right)
        \frac{1}{\sinh{\varphi}}\frac{\partial W}{\partial \varphi}\nonumber \\
        &=&\frac{g^2}{16 W^2 \sinh{\varphi}}\left[
        -8 A' \sin{4\vartheta}\sinh^3{\varphi}\phantom{\frac{e^{-A}}{\ell}}\right.\nonumber
        \\
        & &\left.+ \left(\frac{\kappa e^{-A}}{\ell}\right)\left(8\cos{4\vartheta}\cosh{\varphi}\sinh^3{\varphi}-14\sinh{2\varphi}-\sinh{4\varphi}\right)
        \right]
\end{eqnarray}
with 
\begin{equation}
    W\equiv|\mathcal{W}_{\pm}| = \frac{g}{4\sqrt{2}}\sqrt{35+28 \cosh{2\varphi}+\cosh{4\varphi}-8\cos{4\vartheta}\sinh^4{\varphi}}\, .
\end{equation}
Together with the equation
\begin{equation}
    A'^2-W^2 + \frac{e^{-2A}}{\ell^2}=0,
\end{equation}
we now have the full set of BPS equations for supersymmetric Janus with $N=(1,1)$ supersymmetry on the defect. As in all of the previous cases, it can be directly verified that these equations imply the second-ordered field equations. We also point out that this solution is genuinely new since known supersymmetric Janus solutions within $N=8$ guaged supergravity preserve $(4,4)$, $(0,2)$ and $(0,1)$ supersymmetries on the defect.
\\
\indent In this case, we are not able to find an analytic solution to the above equations. We will instead give an example of numerical solutions as shown in figure \ref{fig4}. In this solution, we have set $g=\frac{1}{\sqrt{2}}$, $\kappa=1$ and $\ell=1$.

%%%%%%%%%%%%%%%%%%%%%%%%%%%%%%%%%%%%%%%%%%%%%%%%%%%%%%%%%%%%%%%%%%%%%%%%%%%%%%%%%%%%%%%%%%%%%%%%%%%%%%%%%%%%%%%%%%%%%%%%%%%%%%%%%%%%%%%%%
\section{Supersymmetric $AdS_4$ black holes}\label{AdS4_BH}
In this section, we consider supersymmetric $AdS_4$ black hole solutions by looking for solutions interpolating between $AdS_4$ and $AdS_2\times \Sigma^2$ geometries. The former is the asymptotic space-time at large distance from the black holes while the latter describes near horizon geometries with $\Sigma^2$ being a two-dimensional Riemann surfaces. In this work, we are only interested in the cases of $\Sigma^2$ being a two-sphere $(S^2)$ and a hyperbolic space $(H^2)$. 
\\
\indent We begin with the metric ansatz
\begin{equation}
    ds^2 = - e^{2 f(r)}dt^2 + dr^2 + e^{ 2 h(r)} (d\theta^2+F^2(\theta)d\phi^2)
\end{equation}
with the function $F(\theta)$ defined by
\begin{equation}
    F(\theta) = 
    \begin{cases}
    \sin \theta, &\,\Sigma^2 = S^2 \\
    \sinh \theta, & \,\Sigma^2 = H^2
    \end{cases}\, .
\end{equation}
It is useful to collect all the non-vanishing components of the spin connection
\begin{eqnarray}
    \omega^{\hat{t}\hat{r}} &=& f' e^{\hat{t}},\qquad    \omega^{\hat{\theta}\hat{r}} = h' e^{\hat{\theta}}, \nonumber \\
    \omega^{\hat{\phi}\hat{r}} &=& h' e^{\hat{\phi}},\qquad
    \omega^{\hat{\theta}\hat{\phi}} = \frac{F'}{F} e^{-h} e^{\hat{\phi}}
\end{eqnarray}
with $F'(\theta)=\frac{dF}{d\theta}$. 
\\
\indent In this case, we need to include gauge fields to the solutions. This is also required by the existence of Killing spinors associated to unbroken supersymmetry. The gauge fields are chosen such that the spin connection, $\omega^{\hat{\theta}\hat{\phi}}$ given above, on $\Sigma^2$ is cancelled. This procedure is called a topological twist. We will consider two possibilities with $SO(2)\times SO(2)$ and $SO(2)$ twists.

\subsection{Solutions with $SO(2)\times SO(2)$ twist}
We first consider $SO(2)\times SO(2)$ twist by turning on $SO(2)\times SO(2)$ gauge fields. We will separately consider magnetic and dyonic solutions.

\subsubsection{Magnetic solutions}
We begin with the ansatz for $SO(2)\times SO(2)$ gauge fields of the form
\begin{equation}
A^{12}=-p_1F'(\theta) d\phi\qquad \textrm{and}\qquad A^{34}=-p_2F'(\theta)d\phi
\end{equation}
with the field strength tensors
\begin{equation}
F^{12}=\kappa p_1F(\theta)d\theta\wedge d\phi\qquad \textrm{and}\qquad F^{34}=\kappa p_2F(\theta)d\theta \wedge d\phi\, .
\end{equation}
We have written $F''(\theta)=-\kappa F(\theta)$ by introducing the parameter $\kappa =1$ and $\kappa =-1$ for $S^2$ and $H^2$, respectively. We also note that $p_1$ and $p_2$ are identified with magnetic charges of the solutions. 
\\
\indent Among the five scalars $\phi^i$, the $SO(2)\times SO(2)$ singlet scalar coincides with the $SO(4)$ singlet $\phi^5=\phi$. With $\phi^i=0$ for $i=1,2,3,4$, it is now straightforward to compute relevant components of the composite connection
\begin{equation}
{{Q_{\hat{\phi}}}^i}_j=-2ge^{-h}\frac{F'(\theta)}{F(\theta)}\left(\begin{array}{ccc}
                                           p_1 \sigma_2& 0_{2\times 2} & 0_{2\times 1} \\
                                            0_{2\times 2} & p_2\sigma_2 & 0_{2\times 1} \\
                                            0_{1\times 2} & 0_{1\times 2} & 0
                                         \end{array}
                                       \right)\, .\label{Q_SO2_SO2}
\end{equation}
From this result, we immediately see that the supersymmetry corresponding to $\epsilon^5$ cannot be preserved since it is not possible to perform a twist along $\epsilon^5$. We then set $\epsilon^5=0$ and split $i,j,\ldots$ indices as $(\hat{i},5)$, $(\hat{j},5)$, $\ldots$.
\\
\indent The twist is implemented by imposing the twist conditions
\begin{equation}
2gp_1=-1\qquad \textrm{and}\qquad 2gp_2=-1\label{SO2_SO2_twist}
\end{equation}
and the following projector on the Killing spinors 
\begin{equation}
\gamma_{\hat{\theta}\hat{\phi}}\epsilon_{\hat{i}}={(i\sigma_2\otimes \mathbb{I}_2)_{\hat{i}}}^{\hat{j}}\epsilon_{\hat{j}}\, .\label{gamma_th_ph_pro}
\end{equation}
The twist conditions imply that $p_2=p_1$ which means the twist is performed by the $SO(2)_{\textrm{diag}}\subset SO(2)\times SO(2)$ gauge field. This is very similar to the solution with a universal twist in pure $N=4$ gauged supergravity studied in \cite{flow_acrossD_bobev}. It is convenient in the analysis of the BPS equations to note that the chirality condition $\gamma_5\epsilon_{\hat{i}}=-\epsilon_{\hat{i}}$ implies that
\begin{equation}
\gamma^{\hat{0}\hat{r}}\epsilon_{\hat{i}}=-i\gamma^{\hat{\theta}\hat{\phi}}\epsilon_{\hat{i}}={(\sigma_2\otimes \mathbb{I}_2)_{\hat{i}}}^{\hat{j}}\epsilon_{\hat{j}}\, .
\end{equation}
\indent To analyze the BPS equations, it is also useful to define the matrix
\begin{equation}
\mc{Z}_{\hat{i}\hat{j}}=\frac{1}{\sqrt{2}}G^+_{\hat{\theta}\hat{\phi}\hat{k}\hat{l}}{C_{\hat{i}\hat{m}}}^{\hat{k}\hat{l}}{(i\sigma_2\otimes \mathbb{I}_2)^{\hat{m}}}_{\hat{j}}
\end{equation}
whose eigenvalues give the ``central charges'' $\mc{Z}_{\hat{i}\hat{j}}=\mc{Z}_{\hat{i}}\delta_{\hat{i}\hat{j}}$ with no summation on $\hat{i}$. For the $SO(2)\times SO(2)$ singlet scalar and $A^{12}_\mu=A^{34}_\mu$ gauge fields, the matrix $\mc{Z}_{\hat{i}\hat{j}}$ is given by
\begin{equation}
\mc{Z}_{\hat{i}\hat{j}}=-\frac{1}{\sqrt{2}}e^{-2h}\kappa p\frac{\sqrt{1-|\phi|^2}}{1+\phi^*}\delta_{\hat{i}\hat{j}}
 \end{equation} 
for $\hat{i},\hat{j}=1,2,3,4$ and $p_1=p_2=p$. We can then identify the central charge as
\begin{equation}
\mc{Z}=-\frac{1}{\sqrt{2}}e^{-2h}\kappa p\frac{\sqrt{1-|\phi|^2}}{1+\phi^*}\, .
 \end{equation} 
We hope the same notation $\phi$ for both scalar $\phi^5$ and the $\Sigma^2$ coordinate will not give rise to any confusion. The two meanings rarely appear together in the same equation.
\\   
\indent With the twist conditions in \eqref{SO2_SO2_twist} and the projector \eqref{gamma_th_ph_pro}, the variations $\delta\psi^i_{\hat{\theta}}$ and $\delta\psi^i_{\hat{\phi}}$ lead to the same BPS equation of the form
\begin{equation}
h'\gamma_{\hat{r}}\epsilon_{\hat{i}}+(\mc{W}+\mc{Z})\epsilon^{\hat{i}}=0\, .
\end{equation}
We again impose the projector \eqref{gamma_r_Pro} and arrive at
\begin{equation}
h'e^{i\Lambda}+\mc{W}+\mc{Z}=0
\end{equation}
which gives
\begin{equation}
h'=\pm |\mc{W}+\mc{Z}|\qquad \textrm{and}\qquad e^{i\Lambda}=\mp \frac{\mc{W}+\mc{Z}}{|\mc{W}+\mc{Z}|}\, .
\end{equation}
Similarly, the condition $\delta\psi^i_{\hat{0}}=0$ gives
\begin{equation}
f'e^{i\Lambda}+\mc{W}-\mc{Z}=0\, .
\end{equation}
Using the phase $e^{i\Lambda}$ from the previous result, we find
\begin{equation}
f'=\pm \frac{(\mc{W}-\mc{Z})(\mc{W}^*+\mc{Z}^*)}{|\mc{W}+\mc{Z}|}\, .
 \end{equation} 
Finally, as in the case of domain walls and Janus solutions, using the phase $e^{i\Lambda}$ in the $\delta\chi^{ijk}=0$ and $\delta\chi=0$ conditions give the BPS equation for the scalar $\phi$. Before giving the explicit form of the resulting BPS equations, we first note that, with $\epsilon^5=0$, $\delta\chi=0$ equation is identically satisfied since this equation has non-vanishing components only along $\epsilon^5$.
\\ 
\indent Since in this case $\mc{W}$ is real, and $\mc{Z}$ is also real for real $\phi $, we firstly consider a simple case of $\phi=\varphi$ for real $\varphi$. It can be readily verified that setting the phase or equivalently the imaginary part of $\phi$ to zero is a consistent truncation. This leads to $e^{i\Lambda}=\mp 1$, and the resulting BPS equations, with the upper sign choice chosen, are given by 
\begin{eqnarray}
\varphi'&=&-\frac{1}{\sqrt{2}}\sqrt{1-\varphi^2}e^{-2h}\left[\varphi(2ge^{2h}-\kappa p)+\kappa p\right],\\
h'&=&\frac{1}{\sqrt{2}}\frac{e^{-2h}}{\sqrt{1-\varphi^2}}\left[2ge^{2h}-\kappa p(1-\varphi)\right],\\
f'&=&\frac{1}{\sqrt{2}}\frac{e^{-2h}}{\sqrt{1-\varphi^2}}\left[2ge^{2h}+\kappa p(1-\varphi)\right].
\end{eqnarray}
\indent In order to find $AdS_2\times \Sigma^2$ fixed points, we impose the conditions $\varphi'=h'=0$ and $f'=\frac{1}{L_{AdS_2}}$. The first two conditions give
\begin{equation}
\varphi=-1 \qquad \textrm{and}\qquad h=\frac{1}{2}\ln \left(\frac{\kappa p}{g}\right),
\end{equation}
and using this result in the last condition gives
\begin{equation}
f'\sim \sqrt{2}g\sqrt{\frac{1-\varphi}{1+\varphi}}\sim \frac{2g}{\sqrt{1+\varphi}}\rightarrow \infty\, .
\end{equation}
Therefore, no $AdS_2\times \Sigma^2$ solutions exist in this case. Note also that the scalar $\varphi$ cannot be truncated out since setting $\varphi=0$ does not satisfy the corresponding flow equation unless $p=0$. We can extend this analysis by including the imaginary part of $\phi$. This will be given in the next subsection. 

\subsubsection{Dyonic solutions}
We now consider dyonic solutions with both magnetic and electric charges. First of all, we review the definition of electric and magnetic charges
\begin{equation}
q_{ij}=\frac{1}{4\pi}\int_{S^2}H_{ij}\qquad \textrm{and}\qquad p^{ij}=\frac{1}{4\pi} \int_{S^2}F^{ij}
\end{equation}
with $H_{ij}$ defined by
\begin{equation}
H_{ij}=\frac{\delta S_{\textrm{gauge}}}{\delta F^{ij}}\, .
\end{equation}
$S_{\textrm{gauge}}$ denotes the gauge field part of the gauged supergravity action.
\\
\indent In the present case, we can rewrite the gauge field part of the Lagrangian by expanding the Lagrangian given in \eqref{L_N5}. The result is
\begin{equation}
\mc{L}_{\textrm{gauge}}=-\frac{1}{4}R_{ij,kl}*F^{ij}\wedge F^{kl}+\frac{1}{4}I_{ij,kl}F^{ij}\wedge F^{kl}
 \end{equation} 
with 
\begin{equation}
R_{ij,kl}=\textrm{Re}(2S^{ij,kl}-\delta^{ik}\delta^{jl})\qquad \textrm{and}\qquad I_{ij,kl}=\textrm{Im}(2S^{ij,kl}-\delta^{ik}\delta^{jl}).
\end{equation}
Using all these results, we can write the ansatz for various components of the gauge field strengths as follows
\begin{eqnarray}
F^{ij}_{\hat{0}\hat{r}}&=&-\frac{1}{2}e^{-2h}R^{ij,kl}\left(\frac{1}{2}I_{kl,mn}\kappa p^{mn}+q_{kl}\right),\\
F_{\hat{\theta}\hat{\phi}}^{ij}&=&\kappa p^{ij}e^{-2h}\, .
\end{eqnarray}
The matrix $R^{ij,kl}$ is the inverse of $R_{ij,kl}$. Note also that $I_{ij,kl}$ is not required to be invertible. Indeed, for real scalars, $I_{ij,kl}$ vanishes identically. In subsequent analysis, we will denote the charges simply by $p^{12}=p_1$, $p^{34}=p_2$, $q_{12}=q_1$ and $q_{34}=q_2$.
\\
\indent  We are now in a position to analyze the BPS conditions. The twist can be performed as in the magnetic case since the components $A^{ij}_{\phi}$ are the same. However, the central charge matrix is now given by 
\begin{equation}
\mc{Z}_{\hat{i}\hat{j}}=\frac{1}{\sqrt{2}}(G^+_{\hat{\theta}\hat{\phi}\hat{k}\hat{l}}-iG^+_{\hat{0}\hat{r}\hat{k}\hat{l}}){C_{\hat{i}\hat{m}}}^{\hat{k}\hat{l}}{(i\sigma_2\otimes \mathbb{I}_2)_{\hat{j}}}^{\hat{m}}\, .
\end{equation} 
All together, we obtain the same form of BPS equations from $\delta\psi^i_{\hat{\theta}}$ and $\delta\psi^i_{\hat{\phi}}$
\begin{equation}
h'=|\mc{W}+\mc{Z}|\qquad \textrm{and}\qquad e^{i\Lambda}=-\frac{(\mc{W}+\mc{Z})}{|\mc{W}+\mc{Z}|}\, .
\end{equation}
\indent With the time component of the composite connection 
\begin{equation}
{Q_{\hat{0}\hat{i}}}^{\hat{j}}=2ige^{-f}A_0{(i\sigma_2\otimes \mathbb{I}_2)_{\hat{i}}}^{\hat{j}},
\end{equation}
the condition $\delta\psi^{\hat{i}}_{\hat{0}}=0$ gives
\begin{eqnarray}
f'\gamma_{\hat{r}}\epsilon_{\hat{i}}+2ige^{-f}A_0\gamma^{\hat{0}}{(i\sigma_2\otimes \mathbb{I}_2)_{\hat{i}}}^{\hat{j}}\epsilon_{\hat{j}}+\mc{W}\epsilon^{\hat{i}}-\mc{Z}\epsilon^{\hat{i}}=0\, .\label{delta_psi_0_dyonic}
\end{eqnarray}
We also recall that the twist conditions require $A^{12}_\mu=A^{34}_\mu=A_\mu$. Since $p_1=p_2=p$, consistency also requires $q_1=q_2=q$.
\\
\indent Using the $\gamma^{\hat{0}\hat{r}}$ and $\gamma^{\hat{r}}$ projectors, we can derive the $\gamma^{\hat{0}}$ projector
\begin{equation}
\gamma^{\hat{0}}\epsilon^{\hat{i}}=-e^{-i\Lambda}{(\sigma_2\otimes \mathbb{I}_2)_{\hat{i}}}^{\hat{j}}\epsilon_{\hat{j}}\, .
\end{equation}
We emphasize here that this is not an independent projector, so the number of unbroken supercharges along the entire flow solutions is still four due to the two projector $\gamma_{\hat{\theta}\hat{\phi}}$ and $\gamma_{\hat{r}}$. Using this result in equation \eqref{delta_psi_0_dyonic} and setting the real and imaginary parts to zero separately, we find the BPS equations
\begin{eqnarray} 
f'&=&-\textrm{Re}\left[e^{-i\Lambda}(\mc{W}-\mc{Z})\right],\\
A_0&=&-\frac{1}{2g}e^{f}\textrm{Im}\left[e^{-i\Lambda}(\mc{W}-\mc{Z})\right]
\end{eqnarray}  
in which the second equation determine the form of the time-component of the gauge fields.   
\\
\indent In the present case, it turns out that using the complex scalar $\phi$ in terms of real and imaginary parts is slightly more convenient. Therefore, we will write
\begin{equation}
\phi=\varphi+i\zeta\, .
\end{equation}      
With the explicit form of the superpotential and the central charge given by
\begin{eqnarray}
\mc{W}&=&\frac{\sqrt{2}g}{\sqrt{1-\varphi^2-\zeta^2}},\\
\textrm{and}\qquad \mc{Z}&=&-\frac{1}{\sqrt{2}}e^{-2h}\frac{\kappa p+2iq+(2iq-\kappa p)(\varphi+i\zeta)}{\sqrt{1-\varphi^2-\zeta^2}}, 
\end{eqnarray}
we find the following BPS equations
\begin{eqnarray}
f'&=&\frac{2\sqrt{2}g[2q(1+\varphi)-\kappa p\zeta]}{\sqrt{1-\varphi^2-\zeta^2}\sqrt{[2q(1+\varphi)-\kappa p\zeta]^2+(2ge^{2h}-\kappa p+\kappa p\varphi+2q\zeta)^2}},\\
h'&=&|\mc{W}+\mc{Z}|=e^{-2h}\sqrt{\frac{(2q+2q\varphi-\kappa p\zeta)^2+(2ge^{2h}-\kappa p+\kappa p\varphi+2q\zeta)^2}{2(1-\varphi^2-\zeta^2)}},\qquad \\
\varphi'&=&(1-\varphi^2-\zeta^2)^2\frac{\pd }{\pd \varphi}|\mc{W}+\mc{Z}|,\\
\zeta'&=&(1-\varphi^2-\zeta^2)^2\frac{\pd }{\pd \zeta}|\mc{W}+\mc{Z}|\, .
\end{eqnarray}
Setting $q=0$, we obtain the BPS equations for magnetic solutions with non-vanishing imaginary part of $\phi$ as previously mentioned. However, even with non-vanishing $\zeta$ and $q$, no $AdS_2\times \Sigma^2$ solutions exist in these equations. Therefore, we conclude that there are no $AdS_4$ black holes with $SO(2)\times SO(2)$ symmetry in $N=5$ gauged supergravity with $SO(5)$ gauge group.

\subsection{Solutions with $SO(2)$ twist}
We now consider $AdS_2\times \Sigma^2$ solutions with $SO(2)$ twist by turning on only $A^{12}_\mu$. The same analysis as in the $SO(2)\times SO(2)$ case can be repeated with $F^{34}_{\mu\nu}=0$ and three $SO(2)$ singlet scalars $\phi^i=\varphi_ie^{i\zeta_i}$, $i=3,4,5$. We will omit some detail to avoid a repetition. The composite connection ${Q_{\hat{\phi}i}}^j$ now has non-vanishing components only for $i,j=\hat{i},\hat{j}=1,2$. In this case, the supersymmetry corresponding to $\epsilon^{3,4,5}$ is broken since it is not possible to perform the twist along these directions. We will accordingly set $\epsilon^{3,4,5}=0$ from now on. With this, $\delta\chi=0$ conditions are identically satisfied as in the $SO(2)\times SO(2)$ case.
\\
\indent As in the case of RG flow solutions, the supersymmetry transformations $\delta\chi^{ijk}$ along $\epsilon^{\hat{i}}$ give rise to the following conditions
\begin{equation}
\zeta_i=\zeta_j+n\pi,\qquad i\neq j, 
\end{equation}
for an interger $n$. In the case of RG flows, there is a possibility to avoid these constraints by setting $\epsilon^{\hat{i}}=0$. However, this is not the case in the present analysis due to the vanishing of $\epsilon^{3,4,5}$ implied by the twist procedure. Therefore, in order to obtain supersymmetric solutions, we need to set
\begin{equation}
\zeta_5=\zeta,\qquad \zeta_4=\zeta+m\pi,\qquad \zeta_3=\zeta+n\pi
\end{equation}  
for $m,n\in \mathbb{Z}$. It turns out that the BPS conditions give
\begin{equation}
\zeta'=0,
\end{equation}
so $\zeta$ is constant and can be set to zero. 
\\
\indent We finally end up with real scalars $\varphi_i$. The variations $\delta\psi^i_{\hat{\phi}}$ and $\delta\psi^i_{\hat{\theta}}$ give
\begin{equation}
e^{i\Lambda}h'=\frac{1}{\sqrt{2}}e^{-2h}\frac{(2ge^{2h}-\kappa p-2iq)}{\sqrt{1-\varphi_3^2-\varphi_4^2-\varphi_5^2}}\, .
\end{equation}
We can immediately see that the condition for $AdS_2\times \Sigma^2$ fixed points to exist, $h'=0$, requires $q=0$. Therefore, the black hole solutions (if exist) must be purely magnetic. Since we are mainly interested in $AdS_4$ black holes, we will set $q=0$ in the following analysis.
\\
\indent For $q=0$, we have real $\mc{W}+\mc{Z}$ which leads to the phase $e^{i\Lambda}=\mp 1$. With the upper sign choice, the resulting BPS equations read
\begin{eqnarray}
f'&=&\frac{2g+\kappa pe^{-2h}}{\sqrt{2(1-\varphi_3^3-\varphi_4^2-\varphi_5^2)}},\\
h'&=&|\mc{W}+\mc{Z}|=\frac{2g-\kappa pe^{-2h}}{\sqrt{2(1-\varphi_3^3-\varphi_4^2-\varphi_5^2)}},\\
\varphi_i'&=&-(1-\varphi_3^2-\varphi_4^2-\varphi_5^2)^2\frac{\pd }{\pd\varphi_i}|\mc{W}+\mc{Z}|\nonumber \\
&=&-\frac{1}{\sqrt{2}}\varphi_i(2g-\kappa pe^{-2h})\sqrt{1-\varphi_3^2-\varphi_4^2-\varphi_5^2},\qquad i=3,4,5\, .
\end{eqnarray}
There is an $AdS_2\times \Sigma^2$ fixed point at 
\begin{equation}
\varphi_i=\varphi_{i(0)},\qquad h=\frac{1}{2}\ln\left[\frac{\kappa p}{2g}\right],\qquad L_{AdS_2}=\frac{\sqrt{1-\varphi^2_{3(0)}-\varphi^2_{4(0)}-\varphi^2_{5(0)}}}{2\sqrt{2}g}\label{AdS2_H2_fixed_point}
\end{equation}
for constant $\varphi_{i(0)}$. By the twist condition $2gp=-1$, we find that the $AdS_2$ fixed point exists only for $\kappa =-1$ giving rise to an $AdS_2\times H^2$ geometry. 
\\
\indent Unlike the previous case with $SO(2)\times SO(2)$ twist, it is possible to truncate all the scalars $\varphi_i$ out resulting in the BPS solution, with $\kappa =-1$,
\begin{eqnarray}
f&=&2\sqrt{2}gr-\frac{1}{2}\ln\left[\frac{e^{2\sqrt{2}gr+C}-p}{2g}\right],\\
\textrm{and}\qquad h&=&\frac{1}{2}\ln\left[\frac{e^{2\sqrt{2}gr+C}-p}{2g}\right]. 
\end{eqnarray}
As $r\rightarrow\infty$, we find $f\sim h\sim \sqrt{2}gr$ which gives asymptotically $AdS_4$ space while for $r\rightarrow -\infty$, the solution becomes 
\begin{equation}
h\sim \frac{1}{2}\ln\left[-\frac{p}{2g}\right]\qquad \textrm{and}\qquad f\sim 2\sqrt{2}gr
\end{equation}
which is the $AdS_2\times H^2$ fixed point. Accordingly, the full solution interpolates between the supersymmetric $AdS_4$ and $AdS_2\times H^2$ geometries. Therefore, this solution describes a black hole in asymptotically $AdS_4$ space with $AdS_2\times H^2$ near horizon geometry. From the holographic point of view, the solution describes twisted compactification of $N=5$ SCFT in three dimensions to superconformal quantum mechanics.
\\
\indent We can also find an analytic solution in the case of non-vanishing scalars $\varphi_i$. We first note that the BPS equations imply, as in the RG flow case, that $\varphi_3=\alpha\varphi_5$ and $\varphi_4=\beta\varphi_5$ for constants $\alpha$ and $\beta$. With $\varphi_5=\varphi$, the solution is given by
\begin{eqnarray}
f&=&\ln\left[2g-p\varphi^2-2g(1+\alpha^2+\beta^2)\varphi^2\right]-\frac{1}{2}\ln\left[1-(1+\alpha^2+\beta^2)\varphi^2\right]\nonumber \\
& &-\ln \varphi,\\
h&=&-\ln \varphi-\frac{1}{2}\ln\left[1-(1+\alpha^2+\beta^2)\varphi^2\right],\\
2\sqrt{2}gr&=&-2\ln\varphi+2\ln \left[1+\sqrt{1-(1+\alpha^2+\beta^2)\varphi^2}\right]\nonumber \\
& & -\sqrt{\frac{p}{p+2g(1+\alpha^2+\beta^2)}}\left\{\ln\left[\frac{8g^2[p+2g(1+\alpha^2+\beta^2)]}{p^3[2g-p\varphi^2-2g(1+\alpha^2+\beta^2)\varphi^2]}\right] \right. \nonumber \\
& &\phantom{\frac{g^2}{p^3}} +\ln \left[(1+\alpha^2+\beta^2)(p\varphi^2-2g+2g(1+\alpha^2+\beta^2)\varphi^2) \phantom{\sqrt{\alpha^2}}\right. \nonumber \\
& &\left.\phantom{\frac{g^2}{p^3}}\left. +2\left(\sqrt{p[p+2g(1+\alpha^2+\beta^2)][1-(1+\alpha^2+\beta^2)\varphi^2]} -p\right)\right]\right\}.\nonumber \\
& &
\end{eqnarray}
From this solution, we can see that as $\varphi\rightarrow 0$, $\ln\varphi\sim -\sqrt{2}gr$ or $\varphi\sim e^{-\sqrt{2}gr}$ and 
\begin{equation}
 f\sim h\sim -\ln\varphi\sim \sqrt{2}gr\, .
\end{equation}
Therefore, the solution is asymptotically $AdS_4$ as in the previous case. 
\\
\indent In order to have a flow to the $AdS_2\times H^2$ fixed point, we require that $\varphi$ flows to the value
\begin{equation}
\varphi_0=\sqrt{\frac{2g}{2g(1+\alpha^2+\beta^2)-p}}
\end{equation} 
which precisely gives $h=\frac{1}{2}\ln \left(-\frac{p}{2g}\right)$ at the end of the flow. As $\varphi\rightarrow \varphi_0$, we find that the above solution gives
\begin{eqnarray}
& &\varphi\sim\varphi_0+C e^{2\sqrt{2}gr\sqrt{\frac{p+2g(1+\alpha^2+\beta^2)}{p}}},\\
\textrm{and}\qquad & &f\sim2\sqrt{2}gr\sqrt{\frac{p+2g(1+\alpha^2+\beta^2)}{p}}\, .
\end{eqnarray}
Therefore, the solution becomes the supersymmetric $AdS_2\times H^2$ fixed point \eqref{AdS2_H2_fixed_point} in the limit $r\rightarrow -\infty$. This solution then describes an $AdS_4$ black hole with $AdS_2\times H^2$ horizon in the presence of a running scalar. 
\\
\indent We end this section by noting that in this case, the flow solution preserves two supercharges due to the $\gamma_{\hat{\theta}\hat{\phi}}$ and $\gamma_{\hat{r}}$ projectors imposed on $\epsilon^{1,2}$. However, the supersymmetry is enhanced to four supercharges at the $AdS_2\times H^2$ horizon.

%%%%%%%%%%%%%%%%%%%%%%%%%%%%%%%%%%%%%%%%%%%%%%%%%%%%%%%%%%%%%%%%%%%%%%%%%%%%%%%%%%%%%%%%%%%%%%%%%%%%%%%%%%%%%%%%%%%%%%%%%%%%%%%%%%%%%%%%%
\section{Conclusions and discussions}\label{conclusion}
In this paper, we have studied supersymmetric solutions of $N=5$ guaged supergravity in four dimensions with $SO(5)$ gauge group. For all scalars vanishing, the gauged supergravity admits an $N=5$ supersymmetric $AdS_4$ vacuum dual to an $N=5$ SCFT in the form of CSM theory in three dimensions. For holographic RG flows describing mass deformations of the $N=5$ SCFT to non-conformal field theories in the IR, we have found analytic solutions preserving $N=5$ supersymmetry, but the $SO(5)$ R-symmetry is broken to an $SO(4)$ subgroup. This is in agreement with the field theory result given in \cite{N5_6_3D_SCFT}. All of the IR singularities are physical by the criterion given in \cite{Gubser_singularity}. Accordingly, these solutions could be useful in the context of the AdS/CFT correspondence regarding the gravity dual of $N=5$ CSM theory in three dimensions. For $SO(3)$ symmetric solutions perserving $N=2$ supersymmetry, we have given numerical flow solutions, but the IR singularities turn out to be unphysical.    
\\
\indent For supersymmetric Janus solutions describing two-dimensional conformal defects within the $N=5$ SCFT, we have studied solutions with $SO(4)$ and $SO(3)$ symmetries and $N=(4,1)$ and $N=(1,1)$ unbroken supersymmetries on the defects, respectively. The former can be found analytically and turns out to be the same as the solutions in $N=8$ and $N=3$ gauged supergravities given in \cite{warner_Janus} and \cite{N3_Janus}. This might suggest some universal property of the solution, and if this is indeed the case, there would be a universal surface defect in the dual three-dimensional SCFTs with $N=3,5,8$ supersymmetries. Further investigation along this direction both in gauged supergravities and dual CSM theories might be worth considering. The $N=(1,1)$ solution with $SO(3)$ symmetry appears to be new and can be only obtained numerically. Both of these solutions could be interesting in the holographic study of strongly coupled $N=5$ SCFT in the presence of conformal defects. 
\\
\indent We have also considered supersymmetric black holes in asymptotically $AdS_4$ space with $SO(2)\times SO(2)$ and $SO(2)$ twists. It turns out that only the case of $SO(2)$ twist leads to a supersymmetric black hole preserving two supercharges with the horizon geometry $AdS_2\times H^2$. In the dual $N=5$ SCFT, the solution describes an RG flow across dimensions from three-dimensional SCFT to superconformal quantum mechanics. This could be used to compute microscopic entropy of the black hole using the formalism initiated in \cite{Zaffaroni_BH_entropy,BH_entropy_benini,BH_entropy_Passias}. It is remarkable that we have found the analytic solution with a running scalar unlike most of the previous analytic solutions that only involve the metric. We accordingly hope our solution would be of particular interest in black hole physics and AdS$_4$/CFT$_3$ correspondence. 
\\
\indent Since the $N=5$ gauged supergravity with $SO(5)$ gauge group considered here is a truncation of the $N=8$ gauged supergravity with $SO(8)$ gauge group, it would be interesting to explicitly find an uplift of these solutions to M-theory using the consistent $S^7$ truncation of the eleven-dimensional supergravity. The uplifted solutions could give rise to a complete holograhic description of $N=5$ CSM theory and possible deformations. In particular, the time component $g_{00}$ of the resulting eleven-dimensional metric can be used to determine whether the aforementioned singular flow solutions are physically acceptable in M-theory by the criterion given in \cite{Maldacena_Nunez_nogo}. 
\\
\indent In this work, we have only considered gauged supergravity with the so-called electric $SO(5)$ gauge group. It could also be interesting to perform a similar study for other gauge groups such as non-compact and non-semisimple ones. In addition, working out the complete embedding tensor formalism of $N=5$ gauged supergravity to incorporate magnetic and dyonic gaugings as initiated in \cite{magetic_gauging_Henning,dyonic_1,dyonic_2}, see also a review \cite{Mario_Physics_Rep}, would be useful in various applications. In particular, the quadratic constraint of $N=5$ theory is generally less stringent than that of the maximal theory. This implies that there are $N=5$ gauged supergravities with certain gauge groups that cannot be obtained from the maximal theory, see a similar analysis for the $N=6$ gauged supergravity in \cite{Henning_twin}. It is then interesting to study these gauged supergravities which would give rise to genuine solutions of $N=5$ gauged supergravity with no $N=8$ counterparts. 

\begin{acknowledgments}
This work is supported by The Thailand Research Fund (TRF) under grant RSA6280022.
\end{acknowledgments}
%%%%%%%%%%%%%%%%%%%%%%%%%%%%%%%%%%%%%%%%%%%%%%%%%%%%%%%%%%%%%%%%%%%%%%%%%%%%%%%%%%%%%%%%%%%%%%%%%%%%%%%%%%%%%%%%%%%%%%%%%%%%%%%%%%%%%%%%%%%%%%%%%%%%%%%%%%%%%%%%

\end{document}